\documentclass[12pt,a4paper]{article} \usepackage[latin1] {inputenc}
\usepackage{amsmath} \usepackage{amsfonts} \usepackage{amssymb}
\usepackage{epsfig}
\usepackage{graphics}
\usepackage{graphicx}
\newcommand{\be}{\begin{equation}}
\newcommand{\ee}{\end{equation}}
\newcommand{\ba}{\begin{eqnarray}}
\newcommand{\ea}{\end{eqnarray}}
\newcommand{\nn}{\nonumber}
\newcommand{\tep}{\tilde{\epsilon}}
\newcommand{\tP}{\tilde{P}}
\begin{document}
\begin{titlepage}

\quad\\
\vspace{1.8cm}
\begin{center}
{\bf\LARGE Quantization of chiral antisymmetric tensor fields}

\vspace{1cm}
C. Wetterich\\
\bigskip
Institut  f\"ur Theoretische Physik\\
Universit\"at Heidelberg\\
Philosophenweg 16, D-69120 Heidelberg\\
\vspace{0.5cm}

\end{center}
\begin{abstract}
Chiral antisymmetric tensor fields can have chiral couplings to quarks and leptons. Their kinetic terms do not mix different representations of the Lorentz symmetry and a local mass term is forbidden by symmetry. The chiral couplings to the fermions are asymptotically free, opening interesting perspectives for a possible solution to the gauge hierarchy problem. We argue that the interacting theory for such fields can be consistently quantized, in contrast to the free theory which is plagued by unstable solutions. We suggest that at the scale where the chiral couplings grow large the electroweak symmetry is spontaneously broken and a mass term for the chiral tensors is generated non-perturbatively. Massive chiral tensors correspond to massive spin one particles that do not have problems of stability. We also propose an equivalent formulation in terms of gauge fields.
\end{abstract}
\end{titlepage}

\section{Introduction}
Chiral antisymmetric tensor fields offer an interesting perspective for an understanding of the scale of weak interactions and a possible solution to the gauge hierarchy problem \cite{1,CWMM}. The chiral couplings to the quarks and leptons are asymptotically free. The scale where they become strong sets a possible scale for the weak interactions, similar to dimensional transmutation in QCD. Indeed, a first solution of the Schwinger-Dyson equations suggests \cite{1} that the strong chiral interactions generate a spontaneous breaking of the electroweak symmetry by a top-antitop condensate \cite{2}. In the absence of spontaneous symmetry breaking a local mass term for the chiral tensors is forbidden by symmetry. Therefore no tuning of parameters is required in order to obtain massless fields, in contrast to a scalar Higgs field.

However, the free theory for chiral antisymmetric tensor fields shows unusual features and the question arises if a unitary quantum theory is possible. In this note we will argue that the interacting theory can indeed be consistently quantized. This requires that the solutions of the quantum field equations, which are derived from the effective action, should be stable and not grow unbounded with time. The energy density for these solutions has to be positive. Furthermore, a consistent quantum field theory has to be covariant under Lorentz transformations. 

While such a consistent picture exists for the interacting theory, the free theory does not admit a bounded Hamiltonian. The free classical solutions show a secular instability. Canonical quantization remains still possible, but there is no consistent theory of free particles for all modes of the non-interacting chiral tensor field. One concludes that the ground state of the interacting theory must be non-perturbative. The free theory is actually on the boundary between the region of stability and instability. Therefore the interactions play a crucial role in the assessment of the stability issue.

At the scale where the asymptotically free chiral couplings between the antisymmetric tensor fields and the quarks and leptons grow large a non-perturbative mass term may be generated \cite{CWMM}. This eliminates the unstable secular solutions. If the mass term is positive all linear classical solutions of the effective field equations remain bounded. The regular solutions  have positive energy density. Massive chiral antisymmetric tensor fields describe massive spin one particles - the chirons. They have no problem of consistency and stability. 

In sect. 2 we introduce the chiral tensors and discuss the symmetries. In lowest order of a derivative expansion and neglecting the interactions the symmetries admit only a unique kinetic term. We summarize the unstable classical solutions of the free theory in sect. 3 and proceed to canonical quantization in sect. 4. The free Hamiltonian operator does not have a spectrum that is bounded from below. However, the instabilities are of a particular type - the unstable secular solutions actually correspond to the borderline between stability and instability. Therefore only the interactions will finally decide if the theory is stable and can be consistently quantized or not. For the interacting theory a quantization by a functional integral is most convenient. We deal with the possible regularizations of this functional integral in sect. 5.

In sect. 6 we turn to the key issue of this paper, namely the stabilization of chiral tensors through interactions. We establish in sect. 7 the general conditions for the effective (full) propagator which are required for stability. We show that a non-perturbative mass term can indeed lead to bounded solutions. In sects. 8-9 we address the issue of a positive energy density. For this purpose we reformulate the model in terms of vector fields and discuss the associated abelian gauge symmetry. In sect. 10 we review the issue of non-perturbative mass generation in a model with chiral couplings to quarks and leptons in the light of this equivalent gauge theory. In two appendices C and D we discuss possible models for chiral tenors in the absence of fermions. We present our conclusions in sect. 11.

\section{Action and symmetries}
The antisymmetric tensors $\beta_{\mu\nu}=-\beta_{\nu\mu}$ describe two irreducible representations $(3,1)$ and $(1,3)$ of the Lorentz group
\begin{equation}\label{1}
\beta^\pm_{\mu\nu}=\frac{1}{2}\beta_{\mu\nu}\pm\frac{i}{4}
\epsilon_{\mu\nu}\ ^{\rho\sigma}\beta_{\rho\sigma}.
\end{equation}
In a realistic  setting with chiral couplings to quarks and leptons the fields $\beta^\pm$ are complex, transform both as doublets with respect to the weak interactions and carry the same nonzero hypercharge. A discrete symmetry $\beta^-\rightarrow -\beta^-$ forbids a local mass term. (The only mass term allowed by the Lorentz symmetry would be of the form $(\beta^+_{\mu\nu})^\dagger\beta^{-\mu\nu}$.) We require our model to be invariant under this ``discrete axial symmetry'' $G_A$. 
Kinetic terms and interactions are allowed by the symmetries.

Our starting point is a functional integral (with Minkowski signature $(-,+,+,+)$)
\begin{eqnarray}\label{2}
Z&=&\int{\cal D}\beta e^{iS_M}~,\nonumber\\
S_M&=&\int d^4 x{\cal L}~,~{\cal L}={\cal L}_{kin}+{\cal L}_{int}.
\end{eqnarray}
The unique invariant kinetic term
\begin{equation}\label{3}
{\cal L}_{kin}=\partial_\mu(\beta^{+\mu\rho})^\dagger\partial_\nu\beta^{+\nu}\ _\rho+\partial_\mu(\beta^{-\mu\rho})^\dagger\partial_\nu\beta^{-\nu}\ _\rho
\end{equation}
has been discussed previously in \cite{3}, \cite{4}, \cite{1}. We emphasize that the kinetic term does not admit an additional abelian gauge symmetry where $\beta_{\mu\nu}$ transforms inhomogeneously as a gauge field. In this sense the chiral tensors should be associated to bosonic matter fields. The term ${\cal L}_{int}$ accounts for the possible interactions with fermions and for quartic self-interactions which will be specified in sect. 6.

The functional integral approach permits an easy control of the Lorentz invariance - this follows simply from the Lorentz invariance of the action $S_M$ and the (postulated) Lorentz invariance of the functional measure $\int {\cal D}\beta$. On the other hand, the unitarity of a quantum theory based on the functional integral (\ref{2}) is less obvious. This requires the identification of a Hamiltonian $\hat{H}$ in the operator formalism which is defined in terms of appropriate operators $\hat{Q}_{\tilde{\alpha}},\hat{P}_{\tilde{\beta}}$ such that the quantum system is equivalent to the functional integral (\ref{2}). Hermiticity of $\hat{H}$ and the boundedness of $\hat{H}$ from below guarantee then a consistent unitary time evolution. We will propose that a Hamiltonian with suitable properties exists for the interacting theory. It is absent for the free theory.

It is convenient to work directly with the unconstrained three component irreducible representations of the Lorentz group $B^\pm_k$. They are related to $\beta^\pm_{\mu\nu}$ by
\begin{eqnarray}\label{5}
\begin{array}{lcl}
\beta^+_{jk}=\epsilon_{jkl}B^+_l&,&\beta^+_{0k}=iB^+_k\quad,\\
\beta^-_{jk}=\epsilon_{jkl}B^-_l&,&\beta^-_{0k}=-iB^-_k.
\end{array}
\end{eqnarray}
In momentum space the kinetic term reads ($\Omega$: four-volume)
\begin{eqnarray}\label{6}
-{\cal L}^{ch}_{\beta,kin}=\Omega^{-1}
\int\frac{d^4q}{(2\pi)^4}
\{(B^{+}_k)^\dagger(q)P_{kl}(q)B^+_l(q)
+(B^-_k)^\dagger(q)P^*_{kl}(q)B^-_l(q)\}
\end{eqnarray}
with $(k,l,j=1\dots 3)$
\begin{equation}\label{7}
P_{kl}=-(q^2_0+q_jq_j)\delta_{kl}+2q_kq_l-2i\epsilon_{klj}q_0q_j.
\end{equation}
The inverse propagator $P_{kl}$ obeys the relations $(q^2=q^\mu q_\mu$)
\begin{eqnarray}\label{No7}
P_{kl}P^*_{lj}&=&(q_kq_k-q^2_0)^2\delta_{mj}=q^4\delta_{mj}\quad,\nonumber\\
 P^*_{kl}(q)&=&P_{lk}(q)
\end{eqnarray}
and $\big(q_\mu=(q_0,\vec{q})~,~\tilde{q}_\mu=(-q_0,~\vec{q})\big)$
\begin{equation}\label{8}
P_{kl}(-q)=P_{kl}(q)~,~P_{kl}(\tilde{q})=P_{lk}(q).
\end{equation}
Clearly $P$ is invertible, $P^{-1}=q^{-4}P^*$ (in a $3\times 3$ matrix notation). 

\section{Free chiral tensors}
\label{free}
Let us first concentrate on the free theory and show where the problems for quantization arise. We begin with the classical theory and start from the Lagrange density (\ref{6}) for the field $B^+_k$ in coordinate space
\begin{eqnarray}\label{S1}
{\cal L}&=&\partial_0B^*_k\partial_0B_k+i\epsilon_{klj}
(\partial_0B^*_k\partial_jB_l-\partial_0B_k\partial_jB^*_l)\nonumber\\
&&+\partial_lB^*_k\partial_lB_k-2\partial_kB^*_k\partial_lB_l.
\end{eqnarray}
(We omit the index $``+''$ - the discussion for $B^-_k$ can be carried out in complete analogy.)
The canonical momenta are
\begin{eqnarray}\label{S2}
\Pi_k&=&\frac{\partial{\cal L}}{\partial(\partial_0B^*_k)}=\partial_0B_k+i\epsilon_{klj}\partial_jB_l
\nonumber\\
\Pi^*_k&=&\frac{\partial{\cal L}}{\partial(\partial_0B_k)}=\partial_0B^*_k-i\epsilon_{klj}\partial_jB^*_l
\end{eqnarray}
and the classical Hamiltonian reads
\begin{eqnarray}\label{S3}
H&=&\int d^3x\{\Pi^*_k\partial_0B_k+\Pi_k\partial_0B^*_k-{\cal L}\}\nonumber\\
&=&\int d^3x\{\Pi^*_k\Pi_k+i\epsilon_{klj}
(\Pi^*_k\partial_lB_j-\Pi_k\partial_lB^*_j)+\partial_kB^*_k\partial_lB_l\}.
\end{eqnarray}
Here we use ${\cal L}=\Pi^*_k\Pi_k-\partial_kB^*_k\partial_lB_l$ (up to a total derivative). We note the unusual mixed term $\sim \Pi^*B$ and the particular form of the space derivatives.

In momentum space the Hamiltonian becomes a sum over independent momentum modes $B_k(\vec{q})$ 
\begin{eqnarray}\label{S4}
H&=&\int\frac{d^3q}{(2\pi)^3}h_{\vec{q}}~,\nonumber\\
h_{\vec{q}}&=&\Pi^*_k(\vec{q})\Pi_k(\vec{q})+q_l\epsilon_{lkj}
\big(\Pi^*_k(\vec{q})B_j(\vec{q})+B^*_j(\vec{q})\Pi_k(\vec{q})\big)\nonumber\\
&&+q_kq_lB^*_k(\vec{q})B_l(\vec{q})
\end{eqnarray}
and we can therefore discuss each momentum mode separately. Without loss of generality we may use $q_1=q_2=0~,~q_3=\bar{q}\geq 0$ such that
\begin{equation}\label{S5}
h=\Pi^*_k\Pi_k+\bar{q}(\Pi^*_1B_2-\Pi^*_2B_1+B^*_2\Pi_1-B^*_1\Pi_2)+\bar{q}^2B^*_3B_3.
\end{equation}
The classical field equations
\begin{equation}\label{S5a}
\dot{B}_k=\frac{\partial h}{\partial\Pi^*_k}~,~\dot{\Pi}_k=-\frac{\partial h}{\partial B^*_k}
\end{equation}
read
\begin{eqnarray}\label{S6}
\dot{B}_3&=&\Pi_3~,~\dot{\Pi}_3=-\bar{q}^2B_3~,\nonumber\\
\dot{B}_1&=&\Pi_1+\bar{q}B_2~,~\dot{\Pi}_1=\bar{q}\Pi_2~,\nonumber\\
\dot{B}_2&=&\Pi_2-\bar{q}B_1~,~\dot{\Pi}_2=-\bar{q}\Pi_1
\end{eqnarray}
or
\begin{eqnarray}\label{S7}
&&\ddot{B}_3+\bar{q}^2B_3=0~,\nonumber\\
&&\ddot{B}_1-2\bar{q}\dot{B}_2-\bar{q}^2B_1=0\nonumber\\
&&\ddot{B}_2+2\bar{q}\dot{B}_1-\bar{q}^2B_2=0.
\end{eqnarray}
For 
\be\label{16AA}
b^\pm_1=(B_1\pm iB_2)/\sqrt{2},~b^\pm_2=(B_1\mp i B_2)/\sqrt{2}~,~ b_3=B_3
\ee
the field equations decouple
\begin{eqnarray}\label{S8}
&&\ddot{b}_1+2i\bar{q}\dot{b}_1-\bar{q}^2b_1=0~,\nonumber\\
&&\ddot{b}_2-2i\bar{q}\dot{b}_2-\bar{q}^2b_2=0.
\end{eqnarray}
Only the longitudinal mode $b_3=B_3$ has standard field equations, whereas the evolution of the transversal modes $b_1,b_2$ is of an unusual type \cite{5A}. As we will see the classical solution for these modes has a ``secular'' instability growing linearly with time. For this reason the modes $b_{1,2}$ have been discarded ad hoc in \cite{5A}. In this note we will discuss later how interactions prevent the secular growth. 

Just as for the Hamiltonian, we could decompose the Lagrangian into a sum of independent modes 
$\big(S=\int dt\int(dq/2\pi)^3{\cal L}_{\vec{q}}\big)$ and choose the basis $b_i(\vec{q})$
\begin{eqnarray}\label{S9}
{\cal L}_{\vec{q}}&=&\partial_0b^*_k\partial_0b_k+\bar{q}^2(b^*_1b_1+b^*_2b_2-b^*_3b_3)\nonumber\\
&&+i\bar{q}\{\partial_0b^*_1b_1-b^*_1\partial_0b_1-\partial_0b^*_2b_2+b^*_2\partial_0b_2\}.
\end{eqnarray}
The canonical momenta $\pi_k$ conjugate to $b_k$ are
\begin{equation}\label{S10}
\pi_3=\dot{b}_3~,~\pi_1=\dot{b}_1+i\bar{q}b_1~,~\pi_2=\dot{b}_2-i\bar{q}b_2
\end{equation}
and it is obvious that the system is unconstrained, i.e. we can solve for $\dot{b}_k$ in terms of $\pi_k$ and $b_k$. The Hamiltonian becomes now a sum over independent modes
\begin{eqnarray}\label{S11}
h&=&\pi^*_1\pi_1-i\bar{q}(\pi^*_1b_1-b^*_1\pi_1)\nonumber\\
&&+\pi^*_2\pi_2+i\bar{q}(\pi^*_2b_2-b^*_2\pi_2)+\pi^*_3\pi_3+\bar{q}^2b^*_3b_3.
\end{eqnarray}
The relation between $\pi$ and $\Pi$ is the same as between $b$ and 
$B,\pi_1=(\Pi_1+i\Pi_2)/\sqrt{2},\pi_2=(\Pi_1-i\Pi_2)/\sqrt{2}$. 

Let us restrict the discussion to $b_1,\pi_1$ and use the familiar notations $b_1=Q,\pi_1=P$,
\begin{equation}\label{S12}
h_1=P^*P-i\bar{q}(P^*Q-Q^*P).
\end{equation}
Of course, the field equation is the same as eq. (\ref{S8})
\begin{eqnarray}\label{S13}
&&\dot{Q}=P-i\bar{q}Q~,~\dot{P}=-i\bar{q}P~,\nonumber\\
&&\ddot{Q}+2i\bar{q}\dot{Q}-\bar{q}^2Q=0.
\end{eqnarray}
It has the general solution
\begin{equation}\label{S14}
Q=(Q_0+P_0t)e^{-i\bar{q}t}~,~P=P_0e^{-i\bar{q}t}
\end{equation}
where we distinguish a ``regular solution'' with a unique frequency for $P_0=0$ from the ``secular solutions'' for which the amplitude of $Q$ grows with time $(P_0\neq 0$). Inserting the solution (\ref{S14}) into $h$ yields the ``on shell Hamiltonian''
\begin{equation}\label{S15}
\bar{h}_1=P^*_0P_0-i\bar{q}(P^*_0Q_0-P_0Q^*_0).
\end{equation}
It vanishes for the regular solution and may take positive or negative values for the secular solutions $\big(Q_0=(Q_{R0}+iQ_{I0}\big)/\sqrt{2},P_0=(P_{R0}+iP_{I0})/\sqrt{2})$
\begin{equation}\label{S16}
\bar{h}_1=\frac{1}{2}\big\{(\bar{q}Q_{R0}-P_{I0})^2-\bar{q}^2Q^2_{R0}+
(\bar{q}Q_{I0}+P_{R0})^2-\bar{q}^2Q^2_{I0}\big\}.
\end{equation}
We conclude that there are classical solutions of the free theory for chiral tensor fields that are not bounded. Also the energy is not bounded from below as can be easily inferred from eq. (\ref{S16}). 

\section{Canonical quantization for chiral tensors}
The Hamiltonian (\ref{S12}) still involves complex degrees of freedom $Q,P$. It therefore describes a system with two real degrees of freedom $Q_\alpha,P_\alpha,\alpha=R,I=1,2$
\begin{eqnarray}\label{T1}
Q&=&(Q_R+iQ_I)/\sqrt{2}~,~P=(P_R+iP_I)/\sqrt{2},\nonumber\\
h_1&=&\frac{1}{2}(P^2_R+P^2_I)+\bar{q}(P_RQ_I-P_IQ_R)\nonumber\\
&=&\frac{1}{2}P_\alpha P_\alpha+\bar{q}\epsilon_{\alpha\beta}P_\alpha Q_\beta.
\end{eqnarray}
At this point we may proceed to canonical quantization. We choose first a normalization adapted to discrete momenta and replace $Q_\alpha,P_\alpha$ and $h$ by operators 
$\hat{Q}_\alpha,\hat{P}_\alpha,\hat{h}$. They obey the commutation relations
\begin{equation}\label{T2}
[\hat{Q}_\alpha,\hat{P}_\beta]=i\delta_{\alpha\beta}~,~
[\hat{Q}_\alpha,\hat{Q}_\beta]=[\hat{P}_\alpha,\hat{P}_\beta]=0.
\end{equation}
In the Heisenberg picture the operators depend on time and the commutation relation (\ref{T2}) holds for arbitrary (equal) time. The time evolution
\begin{equation}\label{T3}
\frac{\partial}{\partial t}\hat{Q}_\alpha=i[\hat{h}_1,\hat{Q}_\alpha]~,~
\frac{\partial}{\partial_t}\hat{P}_\alpha=i[\hat{h}_1,\hat{P}_\alpha]
\end{equation}
reproduces for 
\begin{equation}\label{T4}
\hat{h}_1=\frac{1}{2}\hat{P}_\alpha\hat{P}_\alpha+\bar{q}\epsilon_{\alpha\beta}\hat{P}_\alpha\hat{Q}_\beta
\end{equation}
the same form as the classical equation (\ref{S13}). We observe that there are no constraints acting on the $Q_\alpha,P_\alpha$ such that the canonical quantization (\ref{T2}) is straightforward. The same holds for the modes $b_2$ and $b_3$ (and similar for $B^-_k$) with the only difference concerning the form of the relevant Hamiltonian $\hat{h}_k$. (For $b_2$ one replaces $\rightarrow -q$ and for $b_3$ one has a harmonic oscillator 
$\hat{h}_3=\frac{1}{2}\hat{P}_\alpha\hat{P}_\alpha+\frac{1}{2}q^2\hat{Q}_\alpha\hat{Q}_\alpha$.) We may also switch to the Schr\"odinger picture and introduce a Hilbert space with wave functions $\psi(Q_\alpha)$ or $\psi(P_\alpha)$. As usual, one may define annihilation and creation operators $(\bar{q}=\sqrt{\vec{q}^2})$ as
\begin{equation}\label{T5}
a_\alpha=\frac{1}{\sqrt{2}}\left(\sqrt{\bar{q}}\hat{Q}_\alpha+\frac{i}{\sqrt{\bar{q}}}\hat{P}_\alpha\right)~,~
a^\dagger_\alpha=\frac{1}{\sqrt{2}}(\sqrt{\bar{q}}\hat{Q}_\alpha-\frac{i}{\sqrt{\bar{q}}}\hat{P}_\alpha)~,~
[a_\alpha,a^\dagger_\beta]=\delta_{\alpha\beta},
\end{equation}
and span the Hilbert space from the eigenstates of the occupation numbers 
$\hat{n}_\alpha=a^\dagger_\alpha a_\alpha$. However, the Hamiltonian for the modes $b_{1,2}$ is not a harmonic oscillator. For $b^+_1$ it rather reads
\begin{equation}\label{T6}
\hat{h}_1=\frac{\bar{q}}{2}\big\{a^\dagger_\alpha a_\alpha+a_\alpha a^\dagger_\alpha
-\frac{1}{2}(a_\alpha a_\alpha+a^\dagger_\alpha a^\dagger_\alpha)+2i\epsilon_{\alpha\beta}a^\dagger_\alpha a_\beta\}.
\end{equation}
Therefore the occupation number eigenstates are not eigenstates of $\hat{h}^+_1$. This indicates already that a standard interpretation in terms of free particles will not be possible. 

How do the unbounded secular classical solutions (\ref{S14}) manifest themselves in the quantum theory? We first note that the Hamiltonian $\hat{h}_1$ (\ref{T4}) can be written as a sum of two commuting pieces
\begin{equation}\label{T7}
\hat{h}_1=\hat{h}_P+\hat{h}_M~,~\hat{h}_P=\frac{1}{2}\hat{P}_\alpha\hat{P}_\alpha~,~
\hat{h}_M=\bar{q}\epsilon_{\alpha\beta}\hat{P}_\alpha\hat{Q}_\beta~,~
[\hat{h}_P,\hat{h}_M]=0.
\end{equation}
Let us consider a wave function that is an eigenstate of $\hat{h}_P$, 
\begin{equation}\label{T8}
\hat{h}_P\psi=|P_0|^2\psi
\end{equation}
and investigate the action of $\hat{h}_M$. We may work in a Fourier-type representation where 
\begin{equation}\label{T9}
\hat{P}_1=x~,~\hat{P}_2=y~,~\hat{Q}_1=i\frac{\partial}{\partial x}~,~
\hat{Q}_2=i\frac{\partial}{\partial y}.
\end{equation}
Using polar coordinates
\begin{equation}\label{T10}
r^2=x^2+y^2~,~\varphi=arctg\frac{x}{y}
\end{equation}
one obtains $\hat{h}_M=-i\bar{q}\partial/\partial\varphi$. The stationary wave functions therefore obey
\begin{equation}\label{T11}
\hat{h}_P\psi=\frac{1}{2}r^2\psi=|P_0|^2\psi~,~\hat{h}_M\psi=-i\bar{q}\frac{\partial}{\partial\varphi}\psi=\epsilon_2\psi.
\end{equation}
With $\psi\sim e^{im\varphi},~m$ integer, we obtain $\epsilon_2=m\bar{q}$ and the spectrum of $\hat{h}_1$ reads
\begin{equation}\label{T12}
\hat{h}_1\psi=(|P_0|^2+m\bar{q})\psi.
\end{equation}

Comparing with eq. (\ref{S15}) we see that the regular solution corresponds to $|P_0|^2=0$ and therefore also $m=0~,~\hat{h}_1\psi=0$. For the secular solutions $P_0\neq0$ the spectrum of the free Hamiltonian is not bounded since $m$ can be an arbitrarily large negative integer. Indeed, we may also interprete $\hat{h}_1$ (\ref{T4}) in terms of momentum and angular momentum in the two-dimensional space spanned by $Q_1, Q_2$.  With $\hat{L}=\hat{Q}_1\hat{P}_2-\hat{Q}_2\vec{P}_1$ this yields
\begin{equation}\label{H1}
\hat{h}_1=\frac{1}{2}\hat{\vec{P}}^2-\bar{q}\hat{L}
\end{equation}
and the unboundedness of $\hat{h}_1$ arises from the unboundedness of $\hat{L}$.

We conclude that the free theory for chiral tensors can be quantized but does not lead to a Hamiltonian which is bounded from below. Correspondingly it has unstable solutions. Nevertheless, this instability is of a very particular type. The free theory contains neither ghosts nor tachyons. In a certain sense it is on the boundary between stability and instability. We demonstrate this by two comments.

\begin{itemize}
\item [(i)] Negative energy occurs for $\bar{q}\neq 0$. For $\bar{q}=0$ one finds a bounded Hamiltonian (\ref{H1}). The corresponding classical solution $Q=Q_0+P_0t$ still shows the secular increase with time but the on-shell Hamiltonian $\bar{h}_1=|P_0|^2$  (\ref{S15}) is positive semi definite. For a massive particle we could always go to the rest frame $(\bar{q}=0)$ where one finds positive energy. Positivity of the energy would then be guaranteed in any arbitrary inertial system. In our case only the massless dispersion relation $\omega=\bar{q}$ invalidates the use of the rest frame. We conclude that the boundedness properties may be very different in presence of a mass term, even though the latter may be tiny.
\item[(ii)] The boundary property of the secular solution is well known in classical mechanics. For $m\ddot{x}+\epsilon x=0$
 one finds stable solutions $x\sim e^{i(\omega t +\varphi)},\omega=\sqrt{\epsilon/m}$ for $\epsilon<0$ whereas an instability
 $x\sim x_1 e^{\gamma t}+x_2 e^{-\gamma t},$ $\gamma=\sqrt{-\epsilon/m}$, occurs for $\epsilon<0$. The case $\epsilon<0$
 corresponds to the typical exponential instability in case of ghosts or tachyons. At the boundary between stability and
 instability, i.e. for $\epsilon=0$, one has the secular solution with linear increase in $t,x=x_0+v_0t$. We conclude that
 already a very small perturbation as a tiny positive $\epsilon$ can change from the unstable secular solution to the stable
 solution.
\end{itemize}

We will argue below that in presence of the interactions the Hamiltonian is well behaved and the system becomes stable. In this case one can proceed to canonical quantization similar to eq. (\ref{T2}) without further problems. Of course, the interactions couple the different momentum modes and we have to consider now the properties of the whole Hamiltonian. The canonical quantization rule (\ref{T2}) is easily carried over to operators $\hat{Q}_{k\alpha},\hat{P}_{k\alpha}$ representing the real and complex parts of $B_k$ and $\Pi_k$ (\ref{S2}), respectively, i.e. 
$B_k\rightarrow\frac{1}{\sqrt{2}}(\hat{Q}_{k1}+i\hat{Q}_{k2})$,~ $\Pi_k\rightarrow\frac{1}{\sqrt{2}}(\hat{P}_{k1}+i\hat{P}_{k2})$. (In addition, all operators $\hat{Q}, \hat{P}$ carry a momentum label that we have not written explicitely here.) The commutation relations 
\begin{equation}\label{V1}
[\hat{Q}_{k\alpha}~,~\hat{P}_{l\beta}]=i\delta_{kl}\delta_{\alpha\beta}~,~[\hat{Q}_{k\alpha},\hat{Q}_{l\beta}]=[\hat{P}_{k\alpha},\hat{P}_{l\beta}]=0
\end{equation}
are invariant under space rotations and therefore hold for an arbitrary direction of $\vec{q}$. Operators for different momenta $\vec{q}\neq\vec{q}\ '$ commute. We now switch to the usual continuum normalization and introduce indices $s=(+,-)=1,2$ for $B^+,B^-$ and $u=1,2$ for the two isospin components. This yields the commutation relation
\begin{equation}\label{V2}
[\hat{Q}^{su}_{k\alpha}(\vec{q}),\hat{P}^{tv}_{l\beta}(\vec{q}~')]=i\delta^{st}
\delta^{uv}\delta_{kl}\delta_{\alpha\beta}(2\pi)^3\delta(\vec{q}-\vec{q}~')
\end{equation}
or, in coordinate space
\begin{equation}\label{V3}
\big[\hat{Q}^{su}_{k\alpha}(\vec{x})~,~\hat{P}^{tv}_{l\beta}(\vec{y})\big]=i\delta^{st}\delta^{uv}\delta_{kl}\delta_{\alpha\beta}\delta(\vec{x}-\vec{y}).
\end{equation}
As usual one has $[\hat{Q}_{\tilde{\alpha}},\hat{Q}_{\tilde{\beta}}]=[\hat{P}_{\tilde{\alpha}},\hat{P}_{\tilde{\beta}}]=0$ with $\tilde{\alpha},\tilde{\beta}$ collective indices.

The Hamiltonian becomes (cf. eq. (\ref{S3})) 
\begin{equation}\label{V4}
\hat{H}=\hat{H}_P+\hat{H}_Q+\hat{H}_M+\hat{H}_{int}
\end{equation}
with
\begin{eqnarray}\label{V5}
\hat{H}_P&=&\frac{1}{2}\int\frac{d^3q}{(2\pi)^3}\hat{P}^{su}_{k\alpha}(\vec{q})\hat{P}^{su}_{k\alpha}(\vec{q})~,\nonumber\\
\hat{H}_Q&=&\frac{1}{2}\int\frac{d^3q}{(2\pi)^3}q_kq_l\hat{Q}^{su}_{k\alpha}(\vec{q})\hat{Q}^{su}_{l\alpha}(\vec{q})~,\nonumber\\
\hat{H}_M&=&\int\frac{d^3q}{(2\pi)^3}\kappa\epsilon_{klj}q_k\hat{P}^{su}_{l\alpha}(q)\hat{Q}^{su}_{j\alpha}(\vec{q})
\end{eqnarray}
where $\kappa=1$ for $s=+$ and $\kappa=-1$ for $s=-$. Again, in the Heisenberg picture the operators $\hat{Q}_{\tilde{\alpha}}(t)$ and $\hat{P}_{\tilde{\beta}}(t)$ depend on time and obey at equal time the commutation relation (\ref{V3}). It is easy to check that the time evolution equations 
\begin{equation}\label{V6}
\frac{\partial}{\partial t}\hat{Q}_{\tilde{\alpha}}=i[\hat{H},\hat{Q}_{\tilde{\alpha}}]~,~
\frac{\partial}{\partial t}\hat{P}_{\tilde{\alpha}}=i[\hat{H},\hat{P}_{\tilde{\alpha}}]
\end{equation}
take the same form as the classical field equations (e.g. eq. (\ref{S6}))for the corresponding replacements.

If the interactions can cure the instability nothing prevents us to quantize the system according to eq. (\ref{V2}). What remains to be shown is the consistency of this quantization with the Lorentz symmetry and the correspondence with our definition of the quantum field theory by a functional integral. Reconstructing the functional integral can be done in a standard way. We start with the commutation relation (\ref{V3}) and the Hamiltonian (\ref{V4}). Using the Heisenberg picture for the operators the construction of the functional integral proceeds in the standard way \cite{5}. We require that the Hamiltonian involves at most two powers of $\hat{P}$ and the quadratic term $\hat{H}_P$ is positive definite. Therefore the Gaussian integration over the $p$-variables can be carried out explicitely. The result is indeed a functional integral with the action (\ref{2}), (\ref{3}). 

The direct proof of the Lorentz-symmetry of the quantum system given by the Hamiltonian (\ref{V4}), (\ref{V5}) and the commutation relations (\ref{V2}) or (\ref{V3}) may get rather involved. Fortunately, we can use our findings for an indirect argument. We know already that the time evolution equations in the Heisenberg picture (\ref{V6}) have a Lorentz-covariant form. This follows from the fact that they have the same form as the classical field equations which are derived from an extremum condition for a manifestly Lorentz invariant action. More generally, we have seen that our quantum system can be described by a functional integral which is Lorentz-invariant provided the functional measure preserves the Lorentz symmetry. Indeed, the action is Lorentz-invariant and a Lorentz-invariant measure (or regularization with a Lorentz-invariant continuum limit) means that this symmetry has no anomaly. Then all Greens-functions - propagators and vertices - are Lorentz covariant. This is all we need. Obviously, it is much easier to establish the Lorentz symmetry on the functional integral level than in the Hamiltonian formalism. 

\section{Regularization of the functional integral}
This almost concludes our demonstration that a consistent quantization of interacting chiral antisymmetric tensor fields is possible provided the problem of the unstable solutions can be cured. The only issue that we have not yet addressed concerns the regularization of the functional integral. This issue is similar in some respects to the regularization problem for chiral fermions. Dimensional regularization needs a prescription how to continue $\epsilon_{\mu\nu\rho\sigma}$ to $4-\epsilon$ dimensions. Pauli-Villars regularization is not possible due to the lack of a local mass term consistent with the symmetries. Lattice regularization needs to recover Lorentz symmetry in an appropriate continuum limit. The most straightforward way seems to regularize the Euclidean functional integral which obtains by analytic continuation. In the absence of gauge interactions one may simply employ an ultraviolet cutoff for the momenta, $q^2<\Lambda^2$. 

Also the ERGE-regularization based on exact functional flow equations \cite{ERGE} seems very appropriate. For this purpose it is sufficient to write down an effective infrared cutoff which is consistent with all symmetries. As an IR-cutoff term we may use the structure (\ref{3}) multiplied with $(k^2/q^2)f(q^2/k^2)$ with $f(0)=1$ and $f(q^2\rightarrow\infty)$ decreasing fast, e.g. exponentially $f\sim\exp(-q^2/k^2)$ of $f=(1-q^2/k^2)\theta(k^2-q^2)$. For $q^2\rightarrow 0$ the cutoff acts like a nonlocal mass term, providing an effective mass $\sim k$ to the chiral tensors. The flow describes the change of the effective action as $k$ is lowered towards zero. Since the flow equations are both ultraviolet and infrared finite the regularization has only to specify the initial value of the flow at some high momentum scale $k=\Lambda$.  (Implementing properly the anomalous Ward  identities in presence of a cutoff the ERGE-regularization can also be extended to gauge theories.) 

The analytic continuation to a Euclidean functional integral needs some care. In fact, with respect to the Euclidean rotation group $SO(4)$ the irreducible antisymmetric tensor representations $(3,1)$ and $(1,3)$ are real and not complex conjugate to each other as for the Lorentz group $SO(1,3)$. (With euclidean signature the complex conjugate of the representation $(3,1)$ is equivalent to $(3,1)$. This situation is completely analogous to chiral fermions where for a Minkowski signature the representations $2_L=(2,1)$ and $2_R=(1,2)$ are complex conjugate to each other whereas they are pseudoreal for euclidean signature.) More in detail, the euclidean irreducible antisymmetric tensor representations are given by
\begin{equation}\label{A11}
\beta^{(E)\pm}_{\mu\nu}=\frac{1}{2}\beta^{(E)\pm}_{\mu\nu}\pm \frac{1}{4} \epsilon_{\mu\nu}{^{\rho\sigma}} \beta^{(E)\pm}_{\rho\sigma}.
\end{equation}
This reflects that the analytic continuation from Minkowski to euclidean signature also includes a ``continuation'' of the totally antisymmetric invariant tensor $\epsilon_{\mu\nu\rho\sigma}\rightarrow -i\epsilon_{\mu\nu\rho\sigma}$.  The euclidean invariant kinetic term
\begin{equation}\label{A11a}
{\cal L}^{(E)}_{kin}=\frac{1}{4}(\partial^\rho\beta^{\mu\nu})^\dag\partial_\rho\beta_{\mu\nu}-
(\partial_\mu\beta^{\mu\nu})^\dagger\partial_\rho\beta^\rho{_\nu}
\end{equation}
obtains from eq. (\ref{3}) by multiplying all objects with upper zero-index by $i$, the ones with lower zero-index by $-i$, as well as an overall minus sign. We note that ${\cal L}^{(E)}_{kin}$ is not positive definite, in accordance with our discussion of the unboundedness of the free theory. 

We can again write the kinetic term in a form similar to eq. (\ref{3}) 
\begin{equation}\label{b12a}
{\cal L}^{(E)}_{kin}=-(\partial_\mu\bar\beta^{+\mu\rho}\partial_\nu\beta^{+\nu}{_\rho}+\partial_\mu\bar\beta^{-\mu\rho}\partial_\nu\beta^{-\nu}{_\rho})
\end{equation}
where $\bar\beta^+$ is now related to the complex conjugate of $\beta^-$
\begin{equation}\label{b12b}
\bar\beta^+_{\mu\nu}=(\beta^-_{\mu\nu})^\dagger~,~\bar\beta^-_{\mu\nu}=(\beta^+_{\mu\nu})^\dagger.
\end{equation}
This definition ensures that $\bar\beta^+$ and $\beta^+$ belong to inequivalent representations of $SO(4)$. With the prescription $(\beta^+_{\mu\nu})^\dagger\rightarrow\bar\beta^+_{\mu\nu}=(\beta^-_{\mu\nu})^\dagger$ (and similar for $(\beta^-_{\mu\nu})^\dagger)$ all pieces of the action can be taken directly over to the euclidean signature (with the appropriate overall minus sign). This also holds for the interactions discussed in the next section. 
For the coupling to fermions one has to use the euclidean $\gamma$-matrices $\gamma^0_E=i\gamma^0_M$. 

One may notice that local mass terms $\sim(\beta^+)^*\beta^+$ are now consistent with the gauge symmetries and euclidean $SO(4)$ rotations. Such mass terms being forbidden for Minkowski signature one may wonder if there exist corresponding symmetries for euclidean signature that would be violated by a mass term. In Minkowski space the discrete axial symmetry $G_A$ could be extended to a continuous axial $U(1)_A$ symmetry \cite{CWMM}. In this case we can easily find a corresponding symmetry that forbids local mass terms in euclidean space, namely axial scaling $\beta^+_{\mu\nu}\rightarrow\alpha\beta^+_{\mu\nu},\beta^-_{\mu\nu}\rightarrow\beta^-_{\mu\nu}/\alpha$ with real $\alpha$. The kinetic term and the coupling to fermions are invariant under axial scaling. However, the self interactions of the chiral tensors violate the $U(1)_A$ symmetry in a realistic setting. Finding the analogue of the discrete $G_A$-symmetry for euclidean signature is not easy. Such a symmetry should allow an invariant $(\beta^\dag_{\mu\nu}\beta^{\mu\nu})^2$, while $\beta^\dag_{\mu\nu}\beta^{\mu\nu}$ should not be invariant. In Minkowski space the combination $\beta^\dag_{\mu\nu}\beta^{\mu\nu}$ has odd $G_A$-parity. With euclidean signature no linear transformation in the $24$-component real vector-space spanned by $\beta_{\mu\nu}$ can change the sign of $\beta^\dag_{\mu\nu}\beta^{\mu\nu}$ since this invariant corresponds to the squared length of a vector and is therefore positive definite for all possible transformed vectors.

Despite the difficulty of finding the explicit euclidean analogue of the $G_A$-symmetry no local mass term will appear in the euclidean effective action. Quite generally, the possibility of analytic continuation forbids all terms in the effective action for which the continuation to Minkowski signature would violate the $G_A$-symmetry of the action in Minkowski space. One may easily check explicitely that no local mass term is generated in one loop order. These remarks conclude our general discussion of the quantization of chiral antisymmetric tensor fields.

\section{Stabilization through interactions}
Let us now turn to the central issue of this paper, namely the investigation of the stability problem in presence of interactions. We first add to the Lagrangian a piece describing the chiral interactions of antisymmetric tensors to quarks and leptons
\begin{eqnarray}\label{a}
-{\cal L}_{ch}&=&\bar{u}_R\bar{F}_U\tilde{\beta}_+q_L-\bar{q}_L
\bar{F}^\dagger_U\stackrel{\eqsim}{\beta}_+u_R\nonumber\\
&&+\bar{d}_R\bar{F}_D\bar{\beta}_-q_L-\bar{q}_L\bar{F}^\dagger_D\beta_-d_R\nonumber\\
&&+\bar{e}_R\bar{F}_L\bar{\beta}_-l_L-\bar{l}_L\bar{F}^\dagger_L\beta_-e_R.
\end{eqnarray}
Here we use
\begin{eqnarray}\label{b}
\beta_\pm &=&\frac{1}{2}\beta^\pm_{mn}\sigma^{mn}=\frac{1}{2}\beta_{mn}\sigma^{mn}_\pm 
=\beta_{\pm}\frac{1\pm\gamma^5}{2},\nonumber\\
\bar{\beta}_\pm&=&\frac{1}{2}(\beta^\pm_{mn})^*\sigma^{mn}
=D^{-1}\beta^\dagger_\pm D =\bar{\beta}_\pm \frac{1\mp\gamma^5}{2},\nonumber\\
\tilde{\beta}_+&=&-i\beta^T_+\tau_2~,~ \stackrel{\eqsim}{\beta}_+=i\tau_2\bar{\beta}_+,
\end{eqnarray}
with $\sigma_{\pm}^{mn}=\frac{1}{2}(1\pm\gamma^5)\sigma^{mn},~\sigma^{mn}=\frac{i}{2}
[\gamma^m,\gamma^n]$ and $\bar{\psi}=\psi^\dagger D,~D=\gamma^0,~\gamma^5=-i\gamma^0\gamma^1\gamma^2\gamma^3,~\psi_L=\frac{1}{2}(1+\gamma^5)\psi$. The transposition $\beta^T$ and $\tau_2$ act in weak isospin space, i.e. on the two components of the weak doublet $\beta^+_{mn}$.

In terms of the fields $B^\pm_k$ one has
\begin{eqnarray}\label{c}
\beta_+&=&-2B^+_k\sigma^k_+,\ \beta_-=-2B^-_k\sigma^k_-,\nonumber\\
\bar{\beta}_+&=&-2B^{+*}_k\sigma^k_-~,~\bar{\beta}_-=-2B^{-*}_k\sigma^k_+,
\end{eqnarray}
with $\sigma^k_\pm$ defined in terms of the Pauli matrices
\begin{equation}\label{d}
\sigma^k_+=\left(\begin{array}{cc}
\tau^k&0\\0&0\end{array}\right)\ ,\ \sigma^k_-=\left(\begin{array}{cc}
0&0\\0&\tau^k\end{array}\right).
\end{equation}
The interaction (\ref{a}) is the most general one with dimension four and respecting Lorentz and electroweak gauge symmetries. It automatically obeys the $G_A$-symmetry, and more generally, is invariant under a continuous axial $U(1)_A$-symmetry. The chiral couplings $\bar{F}_{U,D,L}$ are dimensionless $3\times 3$ matrices in generation space. 

The other possible interactions with dimension four are self-interactions for the chiral tensors. The most general quartic interactions consistent with the discrete $G_A$ symmetry can be written as 
\begin{eqnarray}\label{9}
-{\cal L}_{\beta,4}&=&
\frac{\tau_+}{4}\big[(B^+_k)^\dagger B^+_l\big]
\big[(B^+_k)^\dagger B^+_l\big]+(+\rightarrow -)\nonumber\\
&&+\tau_1\big[(B^+_k)^\dagger B^-_k\big]
\big[(B^-_l)^\dagger B^+_l\big]\nonumber\\
&&+\tau_2\big[(B^+_k)^\dagger\vec{\tau}B^-_k\big]
\big[(B^-_l)^\dagger\vec{\tau}B^+_l\big]\nonumber\\
&&+\frac{\tau_3}{4}\big[(B^+_k)^\dagger B^-_k\big]
\big[(B^+_l)^\dagger B^-_l\big]+c.c.\nonumber\\
&&+\frac{\tau_4}{4}\big[(B^+_k)^\dagger B^-_l\big]
\big[(B^+_k)^\dagger B^-_l\big]+c.c.\nonumber\\
&&+\frac{\tau_5}{4}\big[(B^+_k)^\dagger B^-_l\big]
\big[(B^+_l)^\dagger B^-_k\big]+c.c. .
\end{eqnarray}
The couplings $\tau_j$ are dimensionless. We observe that the classical action (\ref{2}), with ${\cal L}_{int}={\cal L}_{ch}+{\cal L}_{\beta,4}$, exhibits no coupling with dimension of mass and is therefore invariant under classical dilatations (scale transformations). 

We now argue that the fluctuation effects due to the interactions may indeed drastically change the issue of stability and could lead to a stable theory. The fluctuation-stabilization of a theory that would be unstable in the absence of interactions is a well known effect. The simplest example is the scalar $\varphi^4$-theory with a classical Lagrange density
\begin{equation}\label{c9a}
-{\cal L}_\phi=\frac{1}{2}\mu^2\varphi^2+\frac\lambda8 \varphi^4+\frac12 \partial^\mu\varphi \partial_\mu\varphi.
\end{equation}
For $\mu^2<0$ the free theory (i.e. neglecting $\lambda$) is unstable with a tachyonic inverse propagator $\sim q^2+\mu^2$. In presence of interactions $(\lambda>0)$ the renormalized mass term in one loop order becomes $\mu^2_R=\mu^2+c\lambda\Lambda^2$ with $\Lambda$ a suitable ultraviolet cutoff (e.g. by a lattice regularization) and $c$ a positive constant. For any finite $\lambda$ there is range of negative $\mu^2$ where the free theory is unstable while the interacting theory is stable due to a positive $\mu^2_R$. No condensate (nonzero expectation value $\langle\varphi\rangle$) occurs in this case. One could have a naive expectation that for small $\lambda$ the interactions play a role only for large field values and therefore the instability of the free theory suggests the occurrence of a condensate. This argument is obviously not valid.

For chiral tensors the issue is more involved since the instability of the free theory is not due to a negative mass term. Unbounded negative eigenvalues of the free Hamiltonian occur now for large (spacelike) momenta $\bar{q}^2$. This time a naive expectation could guess a spontaneous breaking of translation symmetry in case of fluctuation stabilization, due to the dominance of modes with large $\bar{q}^2$. We will argue that this is not the case and suggest that the fluctuation-stabilization of our model rather occurs through the generation of a ``nonlocal mass term''. 

The secular instability of the classical theory arises from solutions with $q^2=0$. In an Euclidean framework we therefore  have to deal with an infrared problem, despite the observation that the most negative eigenvalues of the free Hamiltonian occur for large spacelike momenta $\bar{q}^2$ (for fixed $L$ in eq. (\ref{H1})). We will have to address the question if the pole in the propagator at $q^2=0$ subsists in the interacting theory. One possible way of stabilization is a shift of the location of the pole to a negative nonzero value $-m^2$. We will see that this can cure the instability of the free theory, even though $m^2$ may be tiny as compared to a typical ultraviolet cutoff scale.

In the presence of interactions the investigation of the possible eigenvalues of the Hamiltonian becomes very complicated in the operator formalism. Fortunately, the question of stability can be directly addressed in the functional integral formulation. For this purpose, one needs to compute the effective action by integrating out all quantum fluctuations. Then the quantum field equations obtain from a variation of the effective action. They are exact. An instability of the theory will now show up in the appearance of solutions of the quantum field equations where fluctuations around the assumed ground state grow with time. A first step is the restriction of the effective action to terms quadratic in the fields. If the time evolution of the corresponding solutions remains bounded the linear theory is stable. 

We emphasize that already on the linear level the solutions of the effective field equations may behave very differently from the solutions of the classical field equations. In fact, a bounded Hamiltonian guarantees the stability of small fluctuations around the ground state. However, this is not a necessary condition and linear stability could be realized also for unbounded Hamiltonians - an example are ghosts in the absence of interactions. Inversely, linear stability is not sufficient to guarantee the boundedness of the Hamiltonian. 

In practice, a full proof of the boundedness of the Hamiltonian in presence of interactions with fermions is typically quite difficult. As for most investigations in other models we will be satisfied here with the somewhat weaker criterion of ``nonlinear stability''. We require that all small fluctuations have a bounded time evolution and positive energy as compared to the assumed ground state.

As a first step we integrate out the fermions. This yields a mean field theory (MFT) effective action which only involves the chiral tensors. In particular, the inverse propagator can be computed explicitely in the MFT-approximation. It differs from the classical inverse progator $P_{kl}(q)$ by a wave function renormalization
\begin{equation}\label{c9b}
\tilde{P}^+_{kl}(q)=Z_+(q^2)P_{kl}(q)~,~\tilde{P}^-_{kl}(q)=Z_-(q^2)P^*_{kl}(q).
\end{equation}
Keeping only the chiral coupling of the top quark $\hat{f}_t$ nonzero one finds
\begin{equation}\label{c9c}
Z_+(q^2)=1+\frac{\hat{f}_t^2}{4\pi^2}\ln\frac{\Lambda^2}{q^2},
\end{equation}
where we have chosen a normalization with $Z_+(\Lambda^2)=1$ for some suitably chosen ultraviolet cutoff scale $\Lambda$. Already at this stage we find an important modification of the infrared behavior. Since $Z_+(q^2)$ diverges for $q^2=0$ the chiron propagator has no simple pole at $q^2=0$ anymore.

Furthermore, the fermion fluctuations induce nonlocal quartic interactions between the chiral tensors at the MFT-level. Indeed, a typical fermion loop contribution to a quartic vertex with two external momenta $q$ and two momenta zero is shown in fig. 1. For $q^2\rightarrow 0$ it diverges $\sim~\ln(\Lambda^2/q^2)$. This also demonstrates the importance of the interactions for the stability problem in a drastic way. For any finite amplitude $B$ of a tensor-fluctuation the interaction-contribution to the MFT-field equations grows to infinity as we approach the on-shell condition $q^2\rightarrow 0$. Higher order interactions typically grow even with inverse powers of $q^2$.

\begin{figure}[htb]
\centering
\includegraphics[scale=0.9]{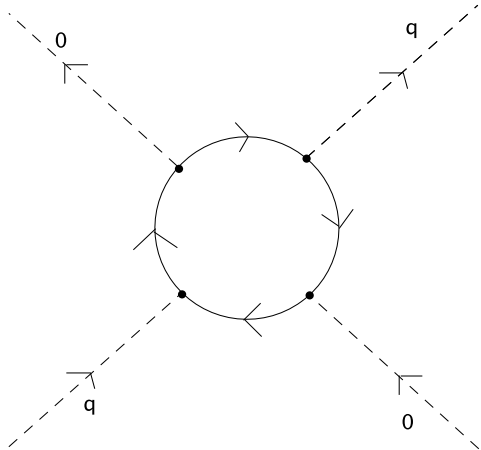}
\caption{Effective quartic vertex for chiral tensors in MFT-approximation.}
\label{quantization-fig1}
\end{figure}

In principle, one could use the MFT-effective action for the quantization of the remaining theory containing only the bosonic chiral tensors. In practice, this is not viable, but a discussion sheds light on the general structure of the problem. 
It is obvious that the Hamiltonian of such a model would be a highly complicated object, with properties quite different from the free Hamiltonian (\ref{H1}). Due to the infrared divergences of the fermion loops it is not possible to consider the MFT-Hamiltonian as a small perturbation of the free Hamiltonian, even though the chiral coupling $\hat{f}_t$ may be small. In the operator language the difference between the Hamiltonians arises since the ground state energy of the fermionic oscillators depends on the state of the chiral tensors, thus inducing an effective Hamiltonian for the chiral tensors.

\section{Conditions for stability}
Taking also the bosonic fluctuations into account it seems likely that the interactions do not diverge on-shell (in contrast to MFT). The stability of small fluctuations is then determined by the properties of the effective propagator. The propagator  will be further modified by the bosonic fluctuation effects and a reliable computation is not easy. We discuss in this section the most general form of the inverse propagator which is consistent with the symmetries. It is specified here for the electrically charged fields $B^+$ and $B^-$ ($\pm$ refers here to chirality, not to electric charge) 
\begin{equation}\label{c9d}
\tilde{P}(q)=\left(\begin{array}{lll}
Z_+(q^2)P_{kl}(q)&,&\mu_1(q^2)\delta_{kl}\\
\mu_2(q^2)\delta_{kl}&,&Z_-(q^2)P^*_{kl}(q)
\end{array}\right).
\end{equation}
Our aim will be to derive criteria for $Z_\pm$ and $\mu^2_j$ for the theory to be stable. We note that nonvanishing $\mu_j(q^2)$ indicate the spontaneous breaking of the $G_A$-symmetry, since such ``mixing terms'' would be forbidden otherwise. If we neglect decays of unstable chiral tensors the wave function renormalizations $Z_\pm$ are real and $\mu_2=\mu^*_1$. 

We first consider $\mu_j=0$ such that the field equation for infinitesimally small fluctuations reads
\begin{equation}\label{YAB}
Z_+(q^2)P_{kl}(q)B_l(t,\vec{q})=0.
\end{equation}
Here we work in momentum space for the spacelike momenta $\vec{q}$, while $q_\mu$ is interpreted as a differential operator 
\begin{equation}\label{d9a}
q_\mu=(-i\frac{\partial}{\partial t},\vec{q})
\end{equation}
such that eq. (\ref{YAB}) is a higher order linear differential equation. We may again use the linear combinations $b_{1,2}$ for which $P_{kl}$ reduces to $(q_0\pm|\vec{q})^2$, cf. eq. (\ref{S8}). For solutions of the type $b_j\sim e^{iq_0t}$ we encounter two possibilities, either $q_0=\pm|\vec{q}|$ or $q^2=-q^2_0+\vec{q}^2=-m^2$ corresponding to a zero of $Z_+(q^2)$. If $Z_+(q^2)$ has at most simple zeros, the secular solutions beyond the exponential ansatz can only arise for $q^2=0$. In particular, as long as $Z_+(q^2\rightarrow0)$ remains finite we find solutions with $q^2=0$ which have the same secular instabilities as for the free theory. A stable theory therefore requires that $Z_+(q^2)$ diverges for $q^2\rightarrow 0$. Furthermore, instabilities occur unless all possible zeros of $Z_+$ at $q^2=-m^2$ have positive real $m^2$. 

Let us look for the conditions of existence of secular solutions in more detail. Defining $(\bar{q}=|\vec{q}|)$
\begin{equation}\label{d9b}
D_\pm=-i\frac{\partial}{\partial t}\pm\bar{q}
\end{equation}
the field equation for $b_1$ reads 
\begin{equation}\label{d9c}
Z_+(-D_+D_-)D^2_+b_1=0.
\end{equation}
We consider the ansatz
\begin{equation}\label{d9d}
b_1(t)=(c_0+c_1t)e^{-i(\bar{q}-\epsilon)t}
\end{equation}
and take the limit $\epsilon\rightarrow 0$ at the end. For $c_1=0$ one has $f(D_+)b_1=f(\epsilon)b_1$, $f(D_-)b_1=f(\epsilon-2\bar{q})b_1$, $Z_+(-D_+D_-)D^2_+ b_1\rightarrow Z_+(2\bar{q}\epsilon)\epsilon^2b_1$. Our ansatz corresponds to a solution provided $\lim\limits_{\epsilon\rightarrow 0} Z_+(2\bar{q}\epsilon)\epsilon^2=0$. This means that solutions with $q^2=0$ exist unless $Z_+(q^2)$ diverges $\sim 1/q^4$ for $q^2\rightarrow 0$ or stronger. Similarly, secular solutions with $c_1\neq0$ exist if $Z_+(q^2)$ diverges less fast than $1/q^2$. In particular, if the interactions could be neglected, the quadratic part in the MFT-effective action would still generate secular solutions. For the particular case $Z_+\sim 1/q^2$ one finds plane wave solutions with $q^2=0$ for $b_1$ and $b_2$ whereas no solutions for $b_3$ with $q^2=0$ exists. 

The situation remains qualitatively similar in presence of spontaneous $G_A$-symmetry breaking for $\mu_1=\mu\neq 0$. The field equations for $B^+$ and $B^-$ are now coupled
\begin{equation}\label{d9e}
\tilde{P}(q)\left(\begin{array}{c}B^+\\B^-\end{array}\right)=0.
\end{equation}
We have to determine the momenta for which at least one eigenvalue of $\tilde{P}(q)$ vanishes. We will choose the momentum frame $q_\mu=(q_0,0,0,\bar{q})$ and the basis $b^\pm_j$ in field space (\ref{16AA}). using $\tilde{b}_1^-=b^-_2,\tilde{b}^-_2=b^-_1, \tilde{b}^-_3=b^-_3$ the inverse propagator is block-diagonal and the field equations read  in the subspaces spanned by $(b^+_k,\tilde{b}^-_k)$ 
\ba\label{b17}
&&\tilde{P}_k(q){b^+_k\choose{\tilde{b}^-_k}}=0,\quad \tilde{P}_3=
\left(\begin{array}{ccc}
Z_+q^2&,&\mu\\\mu^*&,&Z_-q^2\end{array}\right),\nn\\
&&\tilde{P}_1=
\left(\begin{array}{ccc}
-Z_+(q_0+\bar{q})^2&,&\mu\\\mu^*&,&-Z_-(q^0-\bar{q})^2\end{array}\right),\nn\\
&&\tilde{P}_2=\left(\begin{array}{ccc}
-Z_+(q_0-\bar{q})^2&,&\mu\\\mu^*&,&-Z_-(q_0+\bar{q})^2\end{array}\right).
\ea

A sufficient condition for a vanishing eigenvalue of $\tilde{P}_k$ is given by $\det\tilde{P}_k(q)=0$, with
\begin{equation}\label{d9f}
\det\tilde{P}_k(q)=Z_+(q^2)Z_-(q^2)q^4-|\mu(q^2)|^2.
\end{equation}
(We assume here $\mu_2=\mu^*_1.)$
If we exclude the accidental case $\lim\limits_{q^2\rightarrow 0}Z_+Z_-=|\mu_0|^2/q^4$, the r.h.s. does not vanish for  $q^2\rightarrow 0$. Then $\det \tilde{P}_k(q)$ can have zeros only for $q^2\neq 0$. We will call the zeros of $\det\tilde{P}_k$ the regular solutions. If these would be the only possible solutions plane waves with $q^2=0$ would be excluded for all $\mu\neq 0$. Let us denote the location of the regular solutions by $q^2=-m^2_j$. If all $m^2_j$ are real and positive the solutions for linear fluctuations are stable. This situation corresponds to  massive particles with masses $m_j$. Indeed, $\det\tilde{P}_k(q)$ actually only depends on $q^2$ and all small fluctuations obey the standard dispersion relation $q^2_0=\vec{q}^2+m^2_j$. Since $\det\tilde{P}_k(q)$ does not depend on $k$ its zeros are also independent of $k$. Each zero has three eigenvectors, one for every $k$. We will see that they describe massive spin one particles. 

A simple example for stable regular solutions is a nonlocal mass term 
\begin{equation}\label{d9g}
Z_\pm(q^2)=1+\frac{m^2_\pm}{q^2},
\end{equation}
provided $m^2_+m^2_->|\mu|^2, \mu={\rm const.}, m^2_\pm>0$. On the other hand, whenever $Z_+Z_-q^4$ reaches zero for $q^2\rightarrow 0$ (e.g. in case of a finite limit $Z_\pm(q^2\rightarrow 0)$) one finds an instability. Since for real $q^2$ one has $\det\tilde{P}_k(q^2=0)=-|\mu_0|^2<0$ and $\det\tilde{P}_k(q^2\rightarrow +\infty)\rightarrow q^4>0$, continuity implies that there must be a zero at $q^2>0$, corresponding to a tachyon. Finally, we note that it is also possible that $Z_+Z_-q^4$ diverges for $q^2\rightarrow 0$ and no zero eigenvalue of $\det\tilde{P}_k(q)$ exists for any $q_\mu$. Then there are no regular solutions describing propagating particles. 

In general, also additional solutions beyond the regular ones can exist. Indeed, if $Z_\pm$ diverges for $q^2\rightarrow 0$ one can have zero eigenvalues of $\tilde{P}_k$ even if $\det\tilde{P}_k\neq 0$. An example is the previously discussed case $\mu=0,Z_\pm=1+m^2_\pm/q^2$. For $q^2\rightarrow 0$ one finds $\det\tilde{P}_k=m^2_+m^2_-$, while we have found above zero eigenvalues of $\tilde{P}_k$ for $q_0=\pm\bar{q}$, i.e. $q^2=0$. For the zero eigenvalue for $b^+_1$ at $q_0=-\bar{q}$ the second eigenvalue for $b^-_2$ diverges, such that the product of the two eigenvalues remains constant. We will call the zero eigenvalues of $\tilde{P}_k(q)$ which occur for $\det\tilde{P}_k(q)\neq0$ the special solutions. If $Z_\pm(q)$ (and $\mu(q)$) are finite functions for all $q^2\neq0$ the special solutions can only occur for $q^2=0$. For this issue it is therefore sufficient to investigate the behavior of $\tilde{P}_k$ in the limit $q^2\rightarrow 0$. 

Let us consider the non-local wave function renormalization (\ref{d9g}) and $\mu(q^2\rightarrow 0)=\mu_0$. Without loss of generality we take $\mu_0$ real and we exclude the special case $m^2_+m^2_-=\mu^2_0$. With 
\be\label{b18a}
\tilde{P}_3(q^2\rightarrow 0)=
\left(\begin{array}{lll}
m^2_+&,&\mu_0\\ \mu_0&,&m^2_-\end{array}\right)
\ee
it is obvious that no special solutions exist for $b_3$. The situation is different for the transversal modes $b_{1,2}$ where
\ba\label{b18b}
\tilde{P}_1(q^2\rightarrow 0)&=&
\left(\begin{array}{cll}
m^2_+\frac{q_0+\bar{q}}{q_0-\bar{q}}&,&\mu_0\\
\mu_0&,&m^2_-\frac{q_0-\bar{q}}{q_0+\bar{q}}\end{array}\right),\nn\\
\tilde{P}_2(q^2\rightarrow 0)&=&
\left(\begin{array}{cll}
m^2_+\frac{q_0-\bar{q}}{q_0+\bar{q}}&,&\mu_0\\
\mu_0&,&m^2_-\frac{q_0+\bar{q}}{q_0-\bar{q}}\end{array}\right).
\ea
The eigenvalues of $\tilde{P}_1$ are
\be\label{b18c}
\lambda_{1\pm}=\frac{1}{2}\left\{
m^2_+\frac{q_0+\bar{q}}{q_0-\bar{q}}+m^2_-
\frac{q_0-\bar{q}}{q_0+\bar{q}}\pm\left[\left(
m^2_+\frac{q_0+\bar{q}}{q_0-\bar{q}}-m^2_-\frac{q_0-\bar{q}}{q_0+\bar{q}}\right)^2+4\mu^2_0\right]^{1/2}\right\}
\ee
whereas for the eigenvalues of $\tilde{P}_2$ the role of $m^2_+$ and $m^2_-$ is exchanged. 

Let us take for $\lambda_1$ the limit $q_0\rightarrow -\bar{q}$. The first term in the square root diverges and we can expand in powers of $\mu^2_0(q_0+\bar{q})$
\be\label{b18d}
\lambda_{1+}=m^2_+\frac{q_0+\bar{q}}{2q_0}
\left(1-\frac{\mu^2_0}{m^2_+m^2_-}\right)
+0((q_0+\bar q)^2).
\ee
We observe indeed a vanishing eigenvalue for $q_0=-\bar{q}$. The other eigenvalue $\lambda_{1-}$ diverges. The eigenvector to $\lambda_{1+}$ becomes purely $b^+_1$ since mixing with $b^-_2$ vanishes $\sim\mu_0(q_0+\bar{q})/(2q_0m^2_-)$. The role of $b^+_1$ and $b^-_2$ is exchanged for $q_0=\bar{q}$: now $b^-_2$ admits a plane wave with $q^2=0$ whereas the inverse propagator for $b^+_1$ diverges. Also the role of $m^2_+$ and $m^2_-$ are exchanged, according to the limiting behavior for $q_0\rightarrow\bar{q}$
\be\label{b18e}
\lambda_{1-}=m^2_-\frac{q_0-\bar{q}}{2q_0}
\left(1-\frac{\mu^2_0}{m^2_+m^2_-}\right).
\ee
The situation in the sector $b^+_2,b^-_1$ is similar.

In summary, we find plane waves with $q_0=-\bar{q}$ for $b^+_1$ and $b^-_1$, whereas plane waves with $q_0=\bar{q}$ exist for
 $b^+_2$ and $b^-_2$. This is precisely the same situation as for $\mu=0$. Similar to the case $\mu=0$ the special plane wave solutions are
 not allowed anymore if $Z_\pm$ diverges for $q^2\rightarrow 0$ as strong as $q^{-4}$ or stronger. For
 $Z_\pm(q)\sim(q^2)^{-\alpha}$ plane wave solutions with $q^2=0$ exist for the range $0<\alpha<2$. No secular solutions exist for this range. (This differs from $\mu=0$ where secular solutions exist for $\alpha<1$.) The special plane wave solutions for $q^2=0$ are always bounded. If the regular solutions are bounded the linear stability remains maintained in presence of possible special solutions. 

The condition for linear stability can now be formulated in a simple way: it is necessary and sufficient that all zeros of $\det\tilde{P}_k(q)$  occur for negative real values of $q^2$. In particular, this necessitates $Z_+Z_-q^4>|\mu|^4$ for all real values $q^2\geq 0$. Beyond linear stability we will next have to address the question if the regular solutions of eq. (\ref{d9e}) with $q^2=-m^2_j$ and the possible special solutions with $q^2=0$ have a positive energy density. As a convenient way to discuss this question we couple our model to gravity in a standard way and extract the energy momentum tensor by variation with respect to the metric. This allows us to make use of field transformations in the effective action and to investigate the energy density for the transformed fields. 

Before entering this discussion we should stress, however, that massless charged particles are inacceptable phenomenologically even if they are stable. Presumably this implies that no linear solutions with $q^2=0
$ should exist\footnote{As a more remote possibility the modes with $q^2=0$ may not correspond to particles for other reasons.}. This typically requires that the infrared singularities should be even stronger than the one in eq. (\ref{d9g}). For example, an increase of $Z_\pm(q^2)$ in the extreme infrared for $q^2\to0$ of the type $Z_\pm\sim(q^2)^{-3}$ or $Z_\pm\sim q^2/|q^2|^3$ would remove the solutions with $q^2=0$. Regular solutions corresponding to massive spin one particles remain possible in this case. One could also have a wave function renormalization of the type $Z_\pm=1+b/(q^2)^2$ for which no solutions of the field equations exist at all. In this paper we will not investigate the extreme infrared limit and simply assume that the field equations admit only regular solutions. We will mainly use the explicit form (\ref{d9g}) but we emphasize that actually only the behavior of $Z_\pm$ in the vicinity of the poles of the propagator at $q^2=-m^2_j$ is actually needed. We also recall that in the presence of decays the inverse propagator is typically not analytic in the complex $q_0$-plane.

\section{Formulation in terms of gauge fields}
In sect. \ref{positive} we will establish the positivity of the energy density for the regular solutions with a nonlocal mass term (\ref{d9g}). We will find that massive chiral tensors correspond to massive spin one fields. It is therefore natural to look for an equivalent formulation of our model in terms of four-vectors instead of antisymmetric tensors. This reformulation will be introduced in the present section. 

For this purpose we express the chiral tensors in terms of vector fields $V^\pm_\mu$
\begin{equation}\label{AAA}
\beta^\pm_{\mu\nu}=\partial_\mu V^\pm_\nu-\partial_\nu V^\pm_\mu\pm\frac i2 
\epsilon_{\mu\nu}\ ^{\rho\sigma}(\partial_\rho V^\pm_\sigma-\partial_\sigma V^\pm_\rho)
\end{equation}
such that $(\partial^2=\partial_\mu\partial^\mu)$
\begin{equation}\label{AGG}
\partial^\mu\beta^\pm_{\mu\nu}=\partial^2 V^\pm_\nu-\partial_\nu\partial^\mu V^\pm_\mu=\partial^\mu V^\pm_{\mu\nu},
\end{equation}
with field strength
\begin{equation}\label{ABC1}
V^\pm_{\mu\nu}=\partial_\mu V^\pm_\nu-\partial_\nu V^\pm_\mu.
\end{equation}
We will concentrate in this section on wave function renormalizations of the form (\ref{d9g}). Neglecting a total derivative the quadratic term for the chiral tensors can then be written in the form 
\begin{eqnarray}\label{ABC}
-{\cal L}^{ch}_{\beta,kin}&=&-\frac12\Big\{(V^+_{\mu\nu})^\dagger(-\partial^2+m^2_+)V^{+\mu\nu}+(V^-_{\mu\nu})^\dagger
(-\partial^2+m^2_-)V^{-\mu\nu}\nonumber\\
&&+\mu\Big[(V^+_{\mu\nu})^\dagger V^{-\mu\nu}+(V^-_{\mu\nu})^\dagger V^{+\mu\nu}\Big]\Big\}.
\end{eqnarray}
In this formulation the effective action is local in the sense that it contains only a finite number of derivatives. However, the presence of terms with up to four derivatives introduces new unusual features that we will discuss in appendix A. 

For any solution of the linear field equations for $\beta^\pm_{\mu\nu}$ the vector fields obeying eq. (\ref{AAA}) are guaranteed to solve the linear field equations for $V^\pm_\mu$, as derived from eq. (\ref{ABC}). Different solutions for $V^\pm_\mu$ may correspond, however, to one and the same solution for $\beta^\pm_{\mu\nu}$, as can be easily inferred from the relation (\ref{AAA}). Indeed, this relation only involves the field strength $V^\pm_{\mu\nu}$ and $\beta^\pm_{\mu\nu}$ is therefore invariant with respect to the local abelian gauge transformations
\begin{equation}\label{AGF}
V^\pm_\mu\rightarrow V^\pm_\mu+\partial_\mu\lambda^\pm.
\end{equation}
One of the four degrees of $V^+_\mu$ is redundant, since $\beta^+_{\mu\nu}$ or $B^+_k$ describe only three degrees of freedom. In turn, also the action (\ref{ABC}) is gauge invariant, as well as the quartic self interactions for $\beta^\pm_{\mu\nu}$ and the interactions with quarks and leptons. In the latter $\beta^\pm_{\mu\nu}$ gets simply replaced by $2V^\pm_{\mu\nu}$, since the projections $(1\pm\gamma^5)/2$ ensure the correct coupling to $V^\pm_{\mu\nu}\pm\tilde{V}^\pm_{\mu\nu}$ automatically (with the dual field strength $\tilde{V}^\pm_{\mu\nu}$ defined in the standard way). Formulated in terms of $V^\pm_\mu$ the model is invariant under a local $U(1)^8$-symmetry, corresponding to eight real (or four complex) parameters $\lambda^\pm$. We emphasize that concepts of locality and gauge symmetry depend on the basis of fields chosen for the description. The effective action remains local once expressed in terms of $V^\pm_\mu$, even in presence of the nonlocality (\ref{d9g}) in terms of $\beta^\pm_{\mu\nu}$. On the other hand, no trace of the gauge symmetry is visible in the formulation in terms of $\beta_{\mu\nu}$. 

Solutions of the field equations for $V^\pm_\mu$ are guaranteed to correspond to solutions of the field equations for $\beta^\pm_{\mu\nu}$ only if the relation (\ref{AAA}) is invertible, i.e. if $V^\pm_\mu$ can be expressed in terms of $\beta^\pm_{\mu\nu}$. This first requires a gauge fixing and we impose
\begin{equation}\label{AHH}
\partial_\mu V^{\pm\mu}=0.
\end{equation}
With this constraint we may write eq. (\ref{AGG}) as 
\begin{equation}\label{AKA}
V^\pm_\nu=\frac{1}{\partial^2}\partial^\mu\beta^\pm_{\mu\nu}
\end{equation}
provided $\partial^2 V^\pm_\nu\neq 0$. For the regular solutions for $\beta^\pm_{\mu\nu}$, which have $q^2\neq0$, this additional constraint is no problem. 

The behavior of the modes with $q^2=0$ is discussed in detail in appendix A. 
This discussion offers interesting prospects of an alternative interpretation of the chiral tensors, namely as an abelian gauge theory with gauge fields $V^\pm_\mu$. Since the fermions carry no charge with respect to this gauge symmetry the coupling of the gauge fields to the fermions only involves the field strength $V^\pm_{\mu\nu}$. In contrast to the usual gauge theories the gauge fields $V^\pm_\mu$ have dimension zero. Therefore $V^\pm_{\mu\nu}$ has dimension of mass and the chiral couplings to the fermions are dimensionless. The classical kinetic term for $V^+_\mu$ involves four  derivatives, according to eq. (\ref{ABC}) with $m^2_\pm=0, \mu=0$. This accounts for the canonical dimension of $V^\pm_\mu$. The classical action is invariant under dilatation transformations. The ``nonlocal mass'' term breaks this scale invariance and appears here as a local term $m^2_\pm(V^\pm_{\mu\nu})^\dagger V^{+\mu\nu}$.

\section{Positive energy for massive chiral tensors}
\label{positive}
We next turn to the condition of a positive energy density for the solutions of the effective field equations. Since no secular solutions exist anymore we can restrict the discussion to plane waves. We concentrate first on the regular solutions  with four-momentum $q_\mu$ obeying $q^2=-m^2_j$. For all Fourier modes with $q^2<0$, as appropriate for the regular solutions of the linear field equations, we may actually define a rescaled vector field with the canonical dimension of mass \cite{CWMM}
\begin{equation}\label{AAE}
S^\pm_\mu=\frac{1}{\sqrt{\partial^2}}\partial^\nu\beta^\pm_{\nu\mu}=\sqrt{\partial^2} V^\pm_\mu~,~\partial^\mu S^\pm_\mu=0.
\end{equation}
The quadratic term remains local in this formulation
\begin{eqnarray}\label{AEE}
-{\cal L}^{kin}_\beta&=&(S^+_\mu)^\dagger(-\partial^2+m^2_+)S^{+\mu}+(S^-_\mu)^\dagger(-\partial^2+m^2_-)S^{-\mu}
\nonumber\\
&&+\mu[(S^+_\mu)^\dagger S^{-\mu}+(S^-_\mu)^\dagger S^{+\mu}].
\end{eqnarray}
However, the maximal number of derivatives is now two and the gauge fixing, which is implicit in the definition of $S^\pm_\mu$, has eliminated the gauge symmetry in eq. (\ref{AEE}). After diagonalization eq. (\ref{AEE}) describes the standard Lagrangian for massive vector fields, with squared masses
\begin{equation}\label{AAE1}
m^2_j=\frac12\{m^2_++m^2_-\pm \sqrt{(m^2_+-m^2_-)^2+4\mu^2}\}.
\end{equation}

The energy density corresponding to the Lagrangian (\ref{AAE}) is indeed positive for the solutions of the linear field equations for $S^\pm_\mu$, provided $m^2_j>0$. This implies positive energy for the regular solutions of the field equations for $\beta^\pm_{\mu\nu}$. We conclude that in presence of the nonlocal wave function renormalization (\ref{d9g}) the fields $\beta^\pm_{\mu\nu}$ describe stable massive spin one particles. Decays into lighter particles will turn those into resonances, and their status is similar to the $\rho^-$ mesons in QCD.

\section{Mass generation for chiral tensors}
\label{mass}
Possible mechanisms for the generation of ``nonlocal mass terms'' $m^2_\pm$ have been discussed in \cite{CWMM}. We only add here  the perspective of the gauge theory in terms of $V^\pm_\mu$. In the gauge theory version the mass term appears in a local form
\begin{equation}\label{b15d}
-{\cal L}_{m^2}=-\frac12\Big\{m^2_+(V^+_{\mu\nu})^\dagger V^{+\mu\nu}+m^2_-(V^-_{\mu\nu})^\dagger V^{-\mu\nu}\Big\}.
\end{equation}
This resembles the standard kinetic term for gauge theories, but we note the ``wrong'' overall sign. (The opposite sign for ${\cal L}_{m^2}$ would lead to unstable solutions.) Also the dimensions are different from a usual gauge theory due to the different dimension of $V^\pm_\mu$. 
The classical symmetry which forbids ${\cal L}_{m^2}$ is dilatation symmetry. This symmetry is broken by quantum anomalies leading to running couplings. The asymptotically free chiral couplings generate a non-perturbative mass scale $\Lambda_{ch}$, very similar to the non-perturbative scale $\Lambda_{QCD}$ in QCD. It seems plausible that mass terms $m^2_\pm\sim\Lambda^2_{ch}$ can be generated. In particular, this holds in presence of cubic couplings $\tilde{\gamma}$ for $\beta$ or $V$ which are generated by electroweak symmetry breaking \cite{CWMM}. Since these effective couplings have dimension of mass it seems not so surprising that they generate effective contributions $m^2_\pm\sim\tilde{\gamma}^2$, as found by an explicit one loop computation \cite{CWMM}. We finally mention that a non-local mass term for chiral tensors has been proposed  in a different context by Chizhov \cite{4} through the mixing with additional gauge bosons. We show in appendix B that such a mixing cannot lead to a stable theory since the additional vector mesons become ghosts.  

A crucial ingredient for the mass generation and for our arguments in favor of the stabilization of chiral tensor theories is the asymptotic freedom of the chiral couplings to the fermions. One may wonder what happens in theories for chiral tensors
 in the absence of fermions, where the only microscopic interactions are the quartic interactions (\ref{9}). In a first glance  the issue of the generation of a nonlocal mass term seems more obscure. On the other hand, one may ask if the presence of the interactions renders the Hamiltonian bounded from below, at least in presence of an ultraviolet cutoff for the momenta. We address this question in the appendix C. Such a boundedness will require certain positivity criteria for the quartic interactions and we will discuss in the appendix D if they can be met.

\section{Conclusions}
In this paper we have presented arguments that theories with chiral tensor fields are consistent quantum field theories. In this respect a theory with interactions with fermions differs from a free theory. The free theory for chiral tensors has an unbounded Hamiltonian and therefore shows instabilities. However, the free theory lies on the boundary between stable and unstable behavior. The interactions are therefore crucial for an assessment of the stability of the model. We argue that the interactions with the fermions become strong at a characteristic scale $\Lambda_{ch}$ and induce a non-perturbative mass term for the chiral tensors $\sim\Lambda^2_{ch}$. In presence of this mass term the tensor fields describe massive spin one particles and the solutions of the field equations are stable. Since also the Lorentz symmetry is preserved the quantization of interacting chiral tensors is consistent. 

The issue of consistency can also be regarded from a completely different viewpoint. Let us neglect the quartic interactions (\ref{9}) and the gauge bosons mediating the usual electroweak and strong interactions. The model is then quadratic in the chiral tensor fields which may be integrated out. We therefore can formulate our model as an equivalent purely fermionic model with a nonlocal quartic interaction
\ba\label{Z1}
-{\cal L}_{4,F}&=&\{\bar u_RF_U\sigma^{\nu\mu}q_L\}\frac{\partial_\nu\partial_\rho}{\partial^4}
\{\bar q_LF^\dagger_U\sigma^\rho\ _\mu u_R\}\nn\\
&&+\{\bar q_LF^\dagger_D\sigma^{\nu\mu}d_R+\bar l_LF^\dagger_L\sigma^{\nu\mu}e_R\}
\frac{\partial_\nu\partial_\rho}{\partial^4}\nn\\
&& \{\bar d_RF_D\sigma^\rho_\mu q_L+\bar e_RF_L\sigma^\rho_\mu l_L\}.
\ea
We can start as well with the fermionic model (with classical action $S=\int_x({\cal L}_{kin,F}+{\cal L}_{4,F})$ and ${\cal L}_{kin,F}$ the kinetic term for free fermions). It defines a relativistic quantum field theory via a functional integral over fermionic Grassmann fields. This functional integral can be fully regulated - in the infrared by considering antiperiodic Grassmann fields on a torus, and in the ultraviolet by a cutoff respecting the Lorentz and chiral symmetries \cite{9}, \cite{ERGE}. There is little doubt that such a model describes some type of consistent quantum field theory - the only question concerns the nature of its ground state.

Of course, the determination of the ground state for the interaction (\ref{Z1}) is rather intricate. Nevertheless, the enhancement of the attractive interaction in the tensor exchange channel for low $q^2$ strongly suggests the use of partial bosonization via a Hubbard-Stratonovich transformation. This precisely yields our model for chiral tensors with an equivalent action $S=\int_x({\cal L}_{kin,F}+{\cal L}_{ch}+{\cal L}_{kin})$. Since this model for chiral tensors coupled to fermions is strictly equivalent to the purely fermionic model with interaction (\ref{Z1}) it should be considered as a consistent quantum field theory as well. We do not think that the addition of the gauge interactions and the quartic tensor interactions spoils this consistency. 

In summary, the absence of a consistent free theory for chiral tensors only indicates that the ground state is not perturbative.  (Actually, this also holds for the Higgs model of spontaneous symmetry breaking where the free theory with a negative mass term- in the limit of vanishing quartic coupling - is tachyonic.)  Nevertheless, the absence of a consistent free theory has probably prevented so far the systematic investigation of the phenomenological consequences of this interesting class of quantum field theories. We hope that our results open the door for a determination of the properties of the ground state and its possible interesting consequences for our understanding of electroweak symmetry breaking.

\newpage
\noindent
{\bf Acknowledgment:}

\noindent
The author would like to thank A. Hebecker for stimulating discussions.

\section*{Appendix A: Equivalent gauge theory and dual shifts}
\renewcommand{\theequation}{A.\arabic{equation}}
\setcounter{equation}{0}
For a wave function renormalization (\ref{d9g}) the field equations for $V_\mu$ have plane wave solutions with $\partial^2V_\mu=0$. Even though the approximation (\ref{d9g}) is not realistic for $q^2\to 0$ the discussion of this case will shed some light on the properties of the unusual gauge field kinetic term (\ref{ABC}). We first address the general relation between the solutions of the effective field equations for $\beta_{\mu\nu}$ and the ones for $V_\mu$. 
Omitting indices and summations we may write the field equations for $V$ as
\begin{equation}\label{b15a}
\frac{\delta{\cal L}}{\delta V}=\frac{\delta{\cal L}}{\delta\beta}\frac{\delta\beta}{\delta V}=0.
\end{equation}
In linear order the field equations for $V$ and $\beta$ read respectively
\begin{equation}\label{b15b}
\frac{\delta{\cal L}}{\delta V}=A_V V=0~,~
\frac{\delta{\cal L}}{\delta\beta}=A_\beta\beta=0.
\end{equation}
For a linear relation between $\beta$ and $V$ one has 
\begin{equation}\label{b15c}
A_V =A_\beta C_{\beta V}~,~ C_{\beta V}=\frac{\delta\beta}{\delta V}.
\end{equation}

For plane waves these relations hold for every momentum mode separately. Nontrivial solutions for $V$ correspond to zero 
eigenvalues of $A_V$ and therefore $\det A_V=0$. From this we can conclude that $A_\beta$ has zero eigenvalues (for the 
same momentum) only if the relation between $\beta$ and $V$ is invertible. Invertibility requires 
a gauge fixing like eq. (\ref{AHH}), but this a sufficient condition only for $q^2\neq 0$, cf. eq. (\ref{AKA}).  After the gauge fixing only three independent components of $V$ remain and invertibility corresponds to $\det C_{\beta V}\neq 0$. For the modes with $q^2=0$ one finds $\det C_{\beta V}=0$ even after gauge fixing. We want to ensure invertibility also for $q^2=0$ and  impose  for the modes with $q^2=0$ an additional condition for $V$ which eliminates the eigenvectors of the zero eigenvalues of $C_{\beta V}$. 

Let us concentrate on the Fourier  modes $V_\mu(q)$ with $q^2=0$ and denote with $\bar{V}$ the nontrivial solutions of the relation
\begin{equation}\label{ASS}
C_{\beta V}\bar{V}=0.
\end{equation}
A new ``gauge transformation'' $V\rightarrow V+\bar{V}$, with $\bar{V}$ an arbitrary vector in the space of solutions of eq. (\ref{ASS}), will leave $\beta$ invariant. The gauge fixing (\ref{AHH}) $q_\mu V^\mu (q)=0$ is not sufficient to eliminate all $\bar{V}$ and has to be supplemented by an additional constraint for $q^2=0$. Once this is done, the solutions of the field equations derived from the effective action for $V$ are fully equivalent to the solutions of the field equations for $\beta$. We can then use the local form of the effective action (\ref{ABC}). In other words, certain solutions of the field equations derived from eq. (\ref{ABC}) with $\partial^2 V^\pm_\mu=0$ correspond to generalized gauge degrees of freedom  (if $m^2_+m^2_-\neq\mu^2$). They decouple from the physical degrees of freedom, e.g. from the quarks and leptons.

Indeed, the formulation in terms of $V^\pm_\mu$ exhibits an additional symmetry beyond the gauge transformations (\ref{AGF}). We can add to $V^+_\mu$ an arbitrary anti-self-dual field $\delta V^+_\mu$. If $\delta V^+_\mu(x)$ obeys
\be\label{AC18}
\partial_\mu \delta V^+_\nu-\partial_\nu\delta V^+_\mu=-\frac i2 \epsilon_{\mu\nu}\ ^{\rho\sigma}
(\partial_\rho\delta V^+_\sigma-\partial_\sigma\delta V^+_\rho)
\ee
the antisymmetric tensor $\beta^+_{\mu\nu}$ (\ref{AAA}) remains unchanged. Similarly, we have the freedom to add to $V^-_\mu$ an arbitrary self-dual field. One may check explicitely that the action based on eq. (\ref{ABC}) is invariant under these transformations\footnote{The standard kinetic term for gauge fields $\sim\int_x F^*_{\mu\nu}F^{\mu\nu}$ is invariant under transformations obeying (\ref{AC18}) or the corresponding one for anti-self dual variations. However, in QCD this is not a symmetry of the action since the coupling between photons and electrons is not invariant.}. We call the transformations
\be\label{AC19}
V^\pm_\mu\rightarrow V^\pm_\mu+\delta V^\pm_\mu~,~\partial_\mu\delta V^\pm_\nu-\partial_\nu\delta V^\pm_\mu =\mp i\epsilon_{\mu\nu}\ ^{\rho\sigma}\partial_\rho\delta V^\pm_\sigma
\ee
the ``dual shifts'' if $\delta V^\pm_\mu$ cannot be represented as $\partial_\mu\lambda^\pm$. 

Let us consider the case where the dual shifts are plane waves. In momentum space the condition (\ref{AC19}) becomes
\be\label{AC20}
q_\mu\delta V^\pm_\nu-q_\nu\delta V^\pm_\mu=\mp i\epsilon_{\mu\nu}\ ^{\rho\sigma} q_\rho\delta V^\pm_\sigma.
\ee
We first concentrate on the modes with $q^2=0$. For the choice $q_1=q_2=0~,~q_0=q_3=\bar{q}$ the constraint for $\delta V^\pm_\mu(q)$ becomes
\be\label{AC21}
\delta V^\pm_1=\pm i\delta V^\pm_2~,~\delta V^\mp_3=\delta V^\mp_0.
\ee
Both relations pick up an additional minus sign if we choose instead $q_0=-\bar{q}$. For $\delta V^\pm_{1,2}=0$ one has $\delta V^\pm_\mu\sim q_\mu$ and the variation (\ref{AC21}) is therefore an abelian gauge transformation. However, the relation (\ref{AC21}) allows also variations with $\delta V^\pm_{1,2}\neq 0$ - the dual shifts. Without loss of generality we can take the dual shifts as purely transversal, i.e. $\delta V^\pm_{3,0}=0$. 

In contrast, we may consider the modes with $q^2\neq0$, taking $q_1=q_2=0~,~q^2_0\neq q^2_3$. The only solution of eq. (\ref{AC20}) is now $\delta V^\pm_{1,2}=0$. Thus for the modes with $q^2\neq0$ no independent dual shifts exist. This reflects our observation that for $q^2\neq0$ the relation between $V^\pm_\mu$ and $\beta^\mp_{\mu\nu}$ is invertible if the gauge fixing (\ref{AHH}) is imposed. In general, the transformations (\ref{AC19}) are local transformations - we may choose independent $\delta V^\pm_\mu(q)$ for the different momentum modes. However, for plane waves most of these transformations correspond to local gauge transformations. Independent dual shifts exist only for the modes with $q^2=0$.

The dual shifts constitute an additional symmetry group under which the action for $V^\pm_\mu$ must be invariant. This restricts the most general form of the gauge invariant action for $V^\pm_\mu$ to such terms that can also be expressed in terms of $\beta^\pm_{\mu\nu}$. At this point we recall that the dual shifts act differently on $V^+_\mu$ and $V^-_\mu$. This is actually the only memory that $V^+$ and $V^-$ are connected to different Lorentz-representations $\beta^+$ and $\beta^-$. In particular, the dual shift invariance forbids a coupling of $V^-$ to the fermion bilinear $\bar{u}_Rq_L$ in eq. (\ref{a}) - the only allowed couplings involve $\bar{q}_Ld_R$ and $\bar{l}_Le_R$. In turn, $V^+$ can only couple to $\bar{u}_Rq_L$. The dual shift symmetry is obviously a crucial ingredient if our model is formulated in terms of gauge fields $V^\pm_\mu$.

The possibility of gauge fixing for the dual shifts is precisely what is needed in order to eliminate all spurious solutions $\bar{V}$ (\ref{ASS})  and to make the relation between $V$ and $\beta$ invertible also for $q^2=0$. We can then discuss the field equations for $V^\pm$ instead of the ones for $\beta^\pm$. For plane waves the field equation derived from eq. (\ref{AHH}) reads
\be\label{c21a}
(q^2+m^2_\pm)(q^2V^\pm_\mu-q_\mu q_\nu V^{\pm\nu})=-\mu(q^2V^\mp_\mu-q_\mu q_\nu V^{\mp\nu}).
\ee
In a frame with $q_\mu=(q_0,0,0,\bar{q}),\bar{q}>0$ we can write the general solution as
\be\label{c21b}
V^\pm_\mu(q)=c^\pm(q)q_\mu+\hat{V}_\mu(q)~,~\hat{V}_3(q)=\frac{q_0}{\bar{q}}\hat{V}_0(q)
\ee
where $c^\pm(q)$ describe gauge modes and $\hat{V}_\mu$ obeys
\be\label{CA4}
(q^2+m^2_\pm)q^2\hat{V}^\pm_\mu=-\mu q^2\hat{V}^\mp_\mu.
\ee
The solutions with $q^2\neq0$ correspond to the massive spin one particles with masses given by eq. (\ref{AAE1}). Besides that one always finds solutions with $q^2=0$. In the limit $m^2_\pm=0~,~\mu=0$ the solutions with $q^2\neq0$ and $q^2=0$ merge into solutions with $q^2=0$. All solutions have now $q^2=0$ and the double pole in the propagator $\sim 1/q^4$ for $V^\pm_\mu$ explains the presence of the secular solutions discussed in sect. \ref{free}. 

Let us next consider the case $m^2_\pm\neq 0~,~\mu=0$. For $q^2=0$ the propagator has now only a simple pole and no secular
 solutions remain. For the modes with $q^2=0$ there is a residual gauge symmetry which leaves the gauge fixing $q_\mu V^\mu=0$
 invariant. Similar to electromagnetism we can employ the gauge symmetry in order to set $\hat{V}_0=\hat{V}_3=0$. The
 remaining solution for $\hat{V}^\pm_\mu(q)$ is then purely transversal - only $\hat{V}_{1,2}$ differ from zero. The two
 linearly independent solutions can be written as $\hat{V}_2=i\hat{V}_1$ and $\hat{V}_2=-i\hat{V}_1$. For $q_0>0$ and $\hat{V}^+$ the second one can be eliminated by a dual shift transformation (\ref{AC21}) and therefore describes a generalized gauge degree of freedom. The ``physical mode'' obeys $\hat{V}_2=i\hat{V}_1$ and corresponds to the linear combination $\hat{V}^+_1-i\hat{V}^+_2$. Inserting $V^+_0=V^+_3=0$ into eq. (\ref{AAA}) one obtains the nonvanishing components of $\beta^+_{\mu\nu}$
\ba\label{c21c}
\beta^+_{01}&=&i\beta^+_{23}=i(q_0V^+_1-i\bar{q}V^+_2),\nn\\
\beta^+_{02}&=&i\beta^+_{31}=i(q_0V^+_2+i\bar{q}V^+_1),
\ea
and, correspondingly
\be\label{94A}
B^+_1=q_0 V^+_1-i\bar{q}V^+_2~,~B^+_2=q_0V^+_2+i\bar{q}V^+_1~,~B^+_3=0,
\ee
or
\ba\label{c21d}
b^+_1&=&\frac{1}{\sqrt{2}}(q_0-\bar{q})(V^+_1+iV^+_2),\nn\\
b^+_2&=&\frac{1}{\sqrt{2}}(q_0+\bar{q})(V^+_1-iV^+_2).
\ea
For $q_0=\bar{q}$ the physical mode obeys $V^+_2=iV^+_1$ such that $b^+_1=0~,~b^+_2=2\sqrt{2}\bar{q}V^+_1$, whereas for $q_0=-\bar{q}$ the physical mode reads $b^+_1=-2\sqrt{2}\bar{q}V^+_1~,~b^+_2=0$. This reflects the properties of the solutions to the field equations for $b_j$: solutions with $q_0=\bar{q}$ exist only for $b_2$ and solutions with $q_0=-\bar{q}$ only for $b_1$. In the gauge theory formulation this strange behavior is a consequence of the dual shift symmetry. No solution with $q^2=0$ exist for $b_3$ and we recover all features of the discussion after eq. (\ref{d9d}). 

\section*{Appendix B: Mixing of chiral tensors and gauge bosons}
\renewcommand{\theequation}{B.\arabic{equation}}
\setcounter{equation}{0}
It has been suggested that a mass term for chiral tensors  can be induced by coupling them to gauge bosons \cite{4}. We show here that this mechanism leads to tachyons or ghosts in the spin one sector and is therefore not acceptable.

We will demonstrate the problem with complex chiral tensor fields $\beta_{\mu\nu}=\beta^+_{\mu\nu}+\beta^-_{\mu\nu}$ and add complex gauge fields $A_\mu$. (The problem is similar for real fields.) We investigate the effective action in momentum space with 
\begin{eqnarray}\label{E1}
\Gamma&=&\int_q\left\{\frac{q^2}{4}\Pi_{\mu\nu\rho\sigma}(q)\beta^{\mu\nu}(q)^*\beta^{\rho\sigma}(q)\right.\nonumber\\
&&+\frac{\mu}{4}\beta^{\mu\nu}(q)^*\beta_{\mu\nu}(q)\nonumber\\
&&-\tilde{\epsilon}\Big[\beta^{\mu\nu}(q)^*F_{\mu\nu}(q)+\beta^{\mu\nu}(q)F^*_{\mu\nu}(q)\Big]\nonumber\\
&&\left.+\frac{1}{4}F^{\mu\nu}(q)^*F_{\mu\nu}(q)+\frac{1}{2}(m^2 g_{\mu\nu}+\frac{1}{\alpha}q_\mu q_\nu)A^{\mu*}(q)A^\nu(q)\right\}
\end{eqnarray}
with
\begin{eqnarray}\label{E2}
\Pi_{\mu\nu\rho\sigma}(q)&=&\frac{1}{2}(g_{\mu\rho}g_{\nu\sigma}-g_{\mu\sigma}g_{\nu\rho})\nonumber\\
&&-\frac{1}{q^2}(q_\mu q_\rho g_{\nu\sigma}-q_\mu q_\sigma g_{\nu\rho}-q_\nu q_\rho g_{\mu\sigma}+q_\nu q_\sigma g_{\mu\rho})
\end{eqnarray}
obeying
\be\label{E3}
\Pi_{\mu\nu\alpha\beta}\Pi^{\alpha\beta}\ _{\rho\sigma}=\frac{1}{2}(g_{\mu\rho}g_{\nu\sigma}-g_{\mu\sigma}g_{\nu\rho}).
\ee
The coupling $\tilde{\epsilon}$ between gauge bosons and tensors involves the field strength $F_{\mu\nu}=\partial_\mu A_\nu-\partial_\nu A_\mu$ and is consistent with the $U(1)\times U(1)$ gauge symmetry. We have added a gauge fixing term with parameter $\alpha$ and a mass term $m^2$ for the gauge bosons. The latter can be thought of as being generated by the usual Higgs mechanism of spontaneous gauge symmetry breaking by complex scalar fields. The local term 
$\sim\mu\Big\{(\beta^{+\mu\nu})^*\beta^-_{\mu\nu}+(\beta^{-\mu\nu})^*\beta^+_{\mu\nu}\Big\}$ mixes the $\beta^+$ and $\beta^-$ components. It violates the chiral $U(1)_A$ symmetry or the discrete symmetry $G_A$ which forbid local mass terms. In our context it may also be induced by spontaneous symmetry breaking through scalar fields.

In order to investigate the spectrum we first solve the field equations for the gauge fields in presence of 
$\beta_{\mu\nu}(q^2\neq-m^2)$
\begin{equation}\label{E4}
P_{\mu\nu}(q)A^\nu(q)=-2i\tilde{\epsilon}q^\nu\beta_{\nu\mu}(q),
\end{equation}
\be\label{E5}
F_{\mu\nu}(q)=\frac{2\tilde{\epsilon}}{q^2+m^2}(q_\mu q_\rho g_{\nu\sigma}-q_\nu q_\rho g_{\mu\sigma})\beta^{\rho\sigma}(q).
\ee
Insertion into the field equation for $\beta$ yields
\begin{equation}\label{E6}
\left[\frac{q^2}{4}\left(1+\frac{4\tilde{\epsilon}^2}{q^2+m^2}\right)\Pi_{\mu\nu\rho\sigma}+
\left(\frac{\mu}{4}-\frac{q^2\tilde{\epsilon}^2}{q^2+m^2}\right)\frac{1}{2}(g_{\mu\rho}g_{\nu\sigma}-g_{\nu\rho}g_{\mu\sigma})\right]
\beta^{\rho\sigma}(q)=0
\end{equation}
or, in the basis of fields $B^\pm_k$
\begin{eqnarray}\label{E7}
\left(1+\frac{4\tilde{\epsilon}^2}{q^2+m^2}\right)P_{kl}(q)B^+_l(q)+
\left(\mu-\frac{4q^2\tilde{\epsilon}^2}{q^2+m^2}\right)B^-_k(q)=0\nonumber\\
\left(1+\frac{4\tilde{\epsilon}^2}{q^2+m^2}\right)P^*_{kl}(q)B^-_l(q)+\left(\mu-\frac{4q^2\tilde{\epsilon}^2}{q^2+m^2}\right)
B^+_k(q)=0.
\end{eqnarray}
We concentrate on generic momenta for which 
\begin{equation}\label{A2a}
\mu(q^2+m^2)\neq 4q^2\tilde{\epsilon}^2.
\end{equation}
With 
\begin{equation}\label{A3a}
A(q^2)=\frac{4q^2\tilde{\epsilon}^2}{q^2+m^2}
\end{equation}
one finds by inserting one of the equations (\ref{E7}) into the other 
\begin{equation}\label{A3b}
(q^2+\mu)\Big(q^2-\mu+2A(q^2)\Big)B^\pm_k(q)=0.
\end{equation}

Stability requires that all nontrivial plane wave solutions occur for negative or zero $q^2$. One of the modes has $q^2=-\mu$ and we therefore require $\mu \geq0$. The other modes obey
\begin{equation}\label{A3c}
q^2=\mu-2A(q^2)
\end{equation}
which results in a quadratic equation
\begin{equation}\label{E8}
q^4+(m^2-\mu+8\tilde{\epsilon}^2)q^2-\mu m^2=0.
\end{equation}
The solutions are
\begin{equation}\label{E9}
q^2=-\frac{1}{2}\left\{8\tep^2-\mu+m^2\pm
\sqrt{(8\tilde{\epsilon}^2-\mu+m^2)^2+4\mu m^2}
\right\}.
\end{equation}
For $m^2=0$ there is a solution with $q^2=0$ and one with $q^2=-(8\tilde{\epsilon}^2-\mu)$. The latter corresponds to a stable particle if $8\tilde{\epsilon}^2>\mu$ - this is the scenario discussed in \cite{4}.

If we turn on a Higgs mechanism the covariant kinetic term for the scalar will produce a positive $m^2=g^2|\varphi_0|^2$ where $g$ is the gauge coupling and $\varphi_0$ the vacuum expectation value of the scalar. For any positive $m^2$ we find a tachyonic mode with positive $q^2$. This may seem strange since the usual Higgs mechanism produces a positive squared mass for the gauge bosons, as can be seen in the limit $\tilde{\epsilon}=0$ where the gauge boson corresponds to the solution with $q^2=-m^2$ (while the chiral tensor is now tachyonic with $q^2=\mu$.) In order to explain our result we observe that for $m^2=0 ~(\varphi_0=0)$ the gauge symmetry remains unbroken and requires a massless gauge boson, as found here. However, the gauge symmetry does not guarantee that the effective kinetic term for the gauge boson is positive. Due to the mixing the effective kinetic term may turn negative such that the gauge bosons appear as ``ghosts''. Indeed with a negative kinetic term a positive mass term would result in a zero of the inverse propagator at $q^2>0$. This is precisely what happens due to the mixing with the chiral tensors - the gauge bosons are turned into ghosts.

In order to see this easily we ``integrate out'' the chiral tensors, i.e. we solve the field equation for $\beta_{\mu\nu}$ in presence of the gauge fields
\begin{equation}\label{A5a}
\frac{q^2}{4}\Pi_{\mu\nu\rho\sigma}(q)\beta^{\rho\sigma}(q)+\frac{\mu}{4}\beta_{\mu\nu}(q)=\tilde{\epsilon}F_{\mu\nu}(q)
\end{equation}
and insert the solution into the effective action. We are interested in the zero of the inverse propagator at $q^2=0$ (for $m^2=0$) and may therefore use the simplified field equation $\beta_{\mu\nu}=(4\tilde{\epsilon}/\mu)F_{\mu\nu}$. Then the effective action becomes for low momenta
\begin{equation}\label{A5x}
\Gamma=\int_q\left(\frac{1}{4}-\frac{4\tilde{\epsilon}^2}{\mu}\right)F^{\mu\nu^*}(q)F_{\mu\nu}(q).
\end{equation}
For $\tep^2>\mu/16$ the effective kinetic term turns negative. We recall that $\tep^2>\mu/8$ was necessary to avoid a tachyon -  thus the model contains either a tachyon or a ghost or both. In presence of interactions a ghost will lead to instabilities  even in the absence of gauge symmetry breaking $(m^2=0)$. We conclude that this scenario is unacceptable.

We may cast the mixing between the gauge bosons and the chiral tensors into the language of the main part of this paper by ``integrating out'' the gauge bosons. Insertion of the field equation (\ref{E5}) into the effective action yields
\ba\label{A7X}
\Gamma&=&\int_q\left\{\frac{q^2}{4}\left(1+\frac{4\tep^2}{q^2+m^2}\right)\Pi_{\mu\nu\rho\sigma}(q)\beta^{\mu\nu^*}(q)\beta^{\rho\sigma}(q)\right.\nn\\
&&\left.+\left(\frac{\mu}{4}-\frac{q^2\tep^2}{q^2+m^2}\right)\beta^{\mu\nu^*}(q)\beta_{\mu\nu}(q)\right\}.
\ea
For $m^2=0$ we see that the mixing indeed contributes to the nonlocal mass term $\Delta m^2_+=\Delta m^2_-=4\tep^2$, but also to $\Delta\mu=-4\tep^2$. For $\det\tP(q)$ in eq. (\ref{c9a})  one finds
\be\label{A7A}
\det\tP(q)=(q^2+4\tep^2)^2-(\mu-4\tep^2)^2
\ee
with zeros at $q^2=-\mu,q^2=-(8\tep^2-\mu)$. 

For $m^2>0$ an additional solution with $q^2>0$ appears, according to eq. (\ref{E9}). In the limit $m^2\rightarrow 0$ this should turn into the ghost at $q^2=0$. However, this mode is no longer visible in eq. (\ref{A7X}) which has only two modes with positive squared mass if $m^2=0, 8\tep^2>\mu,\mu>0$. The reason for the apparent disappearing of the ghost mode reflects that eq. (\ref{E4}) is no longer invertible if $q^2=0,m^2=0$ and therefore the effective action (\ref{A7X}) is not valid for $q^2+m^2=0$. 

Of course, we can get all these results by investigating directly the zero eigenvalues of the inverse propagator matrix in the space of chiral tensor and gauge fields. We also mention that an investigation of the special case $\mu(q^2+m^2)=4q^2\tep^2$ does not change the situation. 

We conclude that the mixing of chiral tensors and gauge bosons should be a subleading effect and cannot be responsible for the generation of a nonlocal mass term for the chiral tensors. A small mixing between the chiral tensors and the gauge bosons is expected in a realistic model for weak interactions \cite{CWMM}. A different mechanism has to produce nonlocal mass terms of the type $m^2_\pm$ as discussed in the main text. Then a small mixing is acceptable, as discussed in \cite{CWMM}.  In eq. (\ref{A7A}) the nonlocal mass term adds an additional piece $\sim m^2_\pm/4$ in front of $\Pi_{\mu\nu\rho\sigma}$. Now, stability can be achieved in a region of small $\tep^2$ where the kinetic term for the gauge fields (\ref{A5x}) remains positive. More precisely, this holds for $m^2_+=m^2_-$, whereas for $m^2_+\neq m^2_-$ a distinction between $\beta^+_{\mu\nu}$ and $\beta^-_{\mu\nu}$ has to be made. The phenomenology of this mixing, which influences the magnetic moment of the muon, has been addressed in \cite{CWMM}. 

\section*{Appendix C: Stability for interacting chiral \\tensors without fermions?}
\renewcommand{\theequation}{C.\arabic{equation}}
\setcounter{equation}{0}

In this appendix we argue that in presence of appropriate positivity properties of the interactions the Hamiltonian for chiral tensors without fermions becomes bounded from below, such that the model can be quantized consistently.

In the operator picture the quartic polynomial in the fields $B_k$ (\ref{9}) directly translates into a corresponding interaction contribution to the Hamiltonian $H_{int}$ which is quartic in $\hat{Q}$. On an extremely simplified level the crucial features of a possible stabilization can already be seen by adding in eq. (\ref{H1}) to $\hat{h}_1$ an interaction piece $\hat{h}_{int}=\lambda(\hat{Q}_\alpha\hat{Q}_\alpha)^2$. Indeed, a term $\sim(B^*_kB_k)^2$ in ${\cal L}$ will induce an additional term in $h$, i.e. $h_{int}\sim\lambda(b_1^*b_1)^2$ or, on the operator level
\begin{eqnarray}\label{T13}
\hat{h}_{int}&=&\lambda(\hat{Q}_\alpha\hat{Q}_\alpha)^2=\lambda\left(
\frac{\partial^2}{\partial x^2}+\frac{\partial^2}{\partial y^2}\right)^2\nonumber\\
&=&\lambda\left(\frac{\partial^2}{\partial r^2}+\frac{1}{r}\frac{\partial}{\partial r}+
\frac{1}{r^2}\frac{\partial^2}{\partial\varphi^2}\right)^2.
\end{eqnarray}
It is obvious that the Hamiltonian $\hat{h}_P+\hat{h}_M+\hat{h}_{int}$ has a bounded spectrum for $\lambda>0$. Inserting $\partial^2/\partial\varphi^2=-m^2$ one may solve the remaining differential equation for the dependence on $r$. Alternatively, in coordinate space the system $\hat{h}_P+\hat{h}_{int}$ corresponds to an unharmonic two-component oscillator and $\hat{h}_M=-iq\partial/\partial\varphi'=mq$ does not affect the boundedness.

We want to show that $\hat{H}$ is bounded from below for suitable positivity properties of $\hat{H}_{int}$. This requires that $H_{int}(B)$ grows in all directions in field space for $B\rightarrow\infty$. With this positivity restriction on the interaction the pieces $\hat{H}_P,\hat{H}_Q$ and $\hat{H}_{int}$ are all positive definite. The only problem for boundedness may arise from $\hat{H}_M$. Let us consider 
\begin{equation}\label{V7}
\hat{H}'_P=\frac{1}{2}\int\frac{d^3q}{(2\pi)^3}(\hat{P}^{su}_{k\alpha}-\kappa\epsilon_{klj}q_l\hat{Q}^{su}_{j\alpha})
(\hat{P}^{su}_{k\alpha}-\kappa\epsilon_{kl'j'}q_{l'}\hat{Q}^{su}_{j'\alpha})
\end{equation}
which obeys
\begin{eqnarray}\label{V8}
\hat{H}'_P&=&\hat{H}_P+\hat{H}_M-\hat{H}_Q-\hat{H}'_Q~,~
\hat{H}=\hat{H}'_P+2\hat{H}_Q+\hat{H}'_Q+\hat{H}_{int},\nonumber\\
H'_Q&=&-\frac{1}{2}\int\frac{d^3q}{(2\pi)^3}\vec{q}\ ^2\hat{Q}^{su}_{k\alpha}\hat{Q}^{su}_{k\alpha}.
\end{eqnarray}
since $\hat{H}'_P$ and $\hat{H}_Q$ are positive semidefinite operators it is sufficient to show that $\hat{H}_{int}+\hat{H}'_Q$ is bounded from below. For definiteness, let us think about some regularized version, for example on a lattice with large but finite volume, where $\vec{q}^2$ is bounded and the momentum integration is replaced by a discrete sum. The boundedness of $\hat{H}_{int}+\hat{H}'_Q$ is dictated by the interaction term $\hat{H}_{int}$ since this contains terms quartic in $\hat{Q}$. Omitting for a moment the various indices one finds in coordinate space a structure of the type
\begin{equation}\label{V9}
\hat{H}_{int}+\hat{H}'_Q=\int d^3x\{\lambda\hat{Q}^4(x)-\frac{1}{2}\partial_k\hat{Q}(x)\partial_k\hat{Q}(x)\}
\end{equation}
This may be evaluated for $\hat{Q}$-eigenstates $\hat{Q}(x)\psi=B(x)\psi$ such that $\hat{H}_{int}+\hat{H}'_Q$ becomes a functional of $B(x)$. Despite the negative sign of the gradient term this is bounded from below since lattice-derivatives will be of the order of  the inverse lattice distance. Therefore the quartic term always dominates for large enough $|B|$, demonstrating that $\hat{H}_{int}+\hat{H}'_Q$ is indeed bounded from below, implying that also the full Hamiltonian $\hat{H}$ is bounded from below. Adding a suitable constant the energy is therefore positive and the quantum system is well defined if $H_{int}$ meet the positivity criteria.

One may suspect that problems may arise since our argument has involved a regularization and the minimum of $\hat{H}_{int}+\hat{H}'_Q$ is realized for functions $B(x)$ with large gradient terms. The problem of finding the state which minimizes the full Hamiltonian $\hat{H}$ is much more complex, however. 
For $\hat{Q}$-eigenfunctions with large $|\partial_kB|^2$ one also obtains a large positive contribution from $\hat{H}'_P$ which cancels the negative contribution from $\hat{H}'_Q$. Our simple arguments involving a regularization should therefore only be used for the formal proof of the boundedness of $\hat{H}$, not for a search of the state minimizing the energy. As we have learned before, negative contributions to $\hat{H}$ arise only from the transversal modes. For those modes $\hat{H}$ is actually linear in the gradient term (i.e. $\hat{H}_M$ in eq. (\ref{V5})) and the coefficient of the linear term involves $\hat{P}$. The problem of finding the ground state (minimum of $\hat{H}$) amounts to solving the quantum field theory for interacting chiral tensor fields.

\section*{Appendix D: Positivity properties of quartic \\
interactions}
\renewcommand{\theequation}{D.\arabic{equation}}
\setcounter{equation}{0}
In this section we investigate if it is possible to achieve a growing $H_{int}$ for growing $|B|$. We discuss the quartic polynomial (\ref{9}) and investigate of $H_{int}$ can be made positive definite for all $B_k\neq 0$ by restricting the space of allowed couplings $\tau_i$. For example, we may only retain positive couplings $\tau_+>0,\tau_->0$ and set all other quartic interactions to zero. However, the resulting Hamiltonian is not positive definite in all directions in field space. To show this for a one-component tensor it is sufficient to consider  $B^+_k$ and use invariance under rotations in order to achieve for the real and imaginary parts $B_{2R}=B_{3R}=0,~B_{3I}=0$. Insertion into eq. (\ref{9}) yields 
\begin{equation}\label{c10}
H_{int}=\frac{\tau_+}{4}\left\{(B^2_{1R}-B^2_{2I})^2+B^4_{1I}+2B^2_{1I}
(B^2_{1R}+B^2_{2I})\right\}.
\end{equation}
Thus $H_{int}$ is positive semidefinite, but it has two flat directions $B_{2I}=\pm B_{1R},~B_{1I}=0$. We may also add an interaction
\begin{equation}\label{AG9}
-{\cal L}_{\beta,4}=\frac{\hat{\tau}}{32}(\beta^\dagger_{\mu\nu}\beta^{\mu\nu})^2=\frac{\hat{\tau}}{2}\big[(B^+_k)^\dagger 
B^-_k+(B^-_k)^\dagger B^+_k\big]^2
\end{equation}
which corresponds to $\tau_1=\hat{\tau},\tau_3=2\hat{\tau},\tau_j=0$ otherwise. 
For $\hat{\tau}>0$ this is again positive semidefinite, but it does not influence the flat directions of the invariant $\sim\tau_+$ if we set $B^-_k=0$.

We will show next that a flat direction is present for arbitrary polynomials in $B^\pm_k$ that respect the Lorentz symmetry. Thus a local form of $H_{int}$ is at best positive semi-definite. Our argument follows from simple group theory. We first consider invariants that do not involve $B^-$. They have to be built from $n_+$ factors $B^+$ and $n_-$ factors $B^{+*}$. With respect to the Lorentz symmetry $B^+$ transforms as $(3,1)$ such that $(B^+)^{n_+}$ is in some reducible representation $(R_+,1)$. Similarly, $(B^{+*})^{n_-}$ belongs to $(1,R_-)$, and the total invariant transforms as $(R_+,R_-)$. Since this must be the singlet representation $(1,1)$ we can conclude that the product of $n_+$ factors $B^+$ must be itself a singlet, and similarly for the $n_-$ factors of $B^{+*}$. Next, a singlet with respect to the Lorentz symmetry must be a singlet with respect to the subgroup of rotations. Therefore the $n_+$ factors of $B^+$ can be written in terms of the two rotation invariants $B^+_kB^+_k$ and $\epsilon_{klm}B^+_kB^+_lB^+_m$. We next express the first rotation invariant in terms of the linear combinations $b_1,b_2,b_3$ as 
\be\label{c19a}
B^+_{k,a}B^+_{k,b}=b_{1,a}b_{2,b}+b_{2,a}b_{1,b}+b_{3,a}b_{3,b}
\ee
where $a,b$ are additional indices if we have more than one sort of complex $B^+$-fields. It is obvious that for $b_2=b_3=0$ this invariant vanishes for arbitrary $b_1$. Also the invariant $\epsilon_{klm}B^+_{k,a}B^+_{l,b}B^+_{m,c}$ vanishes (even in presence of arbitrary internal indices $a,b,c$) since it involves at least one factor of $B_3=b_3$. We conclude that all possible polynomial invariants constructed only from $B^+$ and $(B^+)^*$ have flat directions, namely $b_2=b_3=0$ with arbitrary $b_1$, and $b_1=b_3=0$ with arbitrary $b_2$. This remains unchanged if we include invariants involving $B^-$, since the flat directions in $B^+$ persist if we put $B^-=0$.

This finding has the far reaching consequence that any effective action which is local in the fields $\beta_{\mu\nu}$, i.e. which admits an expansion in powers of derivatives around constant $\beta_{\mu\nu}$, will necessarily lead to flat directions in field space. The corresponding modes are massless, i.e. the inverse propagator vanishes for $q^2=0$, and furthermore there is no potential in these directions. This situation changes only once nonlocal terms are generated, like the nonlocal mass term $m^2_+$. Due to the nontrivial Lorentz-transformation properties of projectors like $P_{kl}(q)/q^2$ new Lorentz structures are admitted and can lift the degeneracy in the flat directions.

\end{document}